%% file: main_03.tex
\documentclass[review,12pt]{elsarticle}

\usepackage[T1]{fontenc}
\usepackage{lmodern}
\usepackage{graphicx}
\usepackage{booktabs}
\usepackage{multirow}
\usepackage{amsmath,amssymb,amsfonts,mathtools}
\usepackage{amsthm}
\usepackage[mathscr]{euscript}
\usepackage{bm}
\usepackage{braket}
\usepackage{empheq}
\usepackage{bbm}
\usepackage[version=4]{mhchem}
\usepackage{siunitx}
\usepackage{caption}
\usepackage{subcaption}
\usepackage{algorithm}
\usepackage{algorithmicx}
\usepackage{algpseudocode}
\usepackage{listings}
\usepackage[inline]{enumitem}
\usepackage{comment}
\usepackage{textcomp}
\usepackage{float}
\usepackage{lineno}
\usepackage{tabularx}
\usepackage{longtable}
\usepackage{setspace}
\usepackage{array}

\usepackage{hyperref}

\AtBeginDocument{%
  \hypersetup{
    hidelinks,
    pdftitle={
      Statistical admissibility and long-wavelength structural convergence
      determine representative support in granular materials
    },
    pdfauthor={
      Christian Tantardini, Fernando Alonso-Marroquin,
      Abdullah Alqubalee, Eduardo Garzanti
    }
  }%
}

\modulolinenumbers[1]

\biboptions{sort&compress}

\begin{document}
\begin{frontmatter}

\title{Statistical admissibility and long-wavelength structural convergence determine representative support in granular materials}

\author[cipr]{Christian Tantardini\corref{cor1}\fnref{equal}}
\ead{christiantantardini@ymail.com}
\author[eth]{Fernando Alonso-Marroqu\'in\fnref{equal}}
\author[cipr]{Abdullah Alqubalee}
\author[milan]{Eduardo Garzanti\corref{cor1}}
\ead{eduardo.garzanti@unimib.it}

\cortext[cor1]{Corresponding author.}
\fntext[equal]{These authors contributed equally to this work.}

\affiliation[eth]{organization={Department of Computational Physics for Engineering Materials, ETH Zurich},
                 city={Zurich}, postcode={8092}, country={Switzerland}}
\affiliation[cipr]{organization={Center for Integrative Petroleum Research, King Fahd University of Petroleum and Minerals},
                  city={Dhahran}, postcode={31261}, country={Saudi Arabia}}
\affiliation[milan]{organization={Dipartimento di Scienze dell'Ambiente e della Terra, Universit\`a degli Studi di Milano-Bicocca},
                   addressline={Piazza della Scienza 4}, city={Milan}, postcode={20126}, country={Italy}}

\begin{abstract}
Representative elementary areas and volumes provide the scale bridge between
resolved granular microstructure and continuum-scale material properties, yet
they are still commonly selected from the apparent stabilization of a single
scalar observable. Such plateaus can be false indicators of
representativeness when statistically incompatible regions are averaged
together or when long-wavelength structural correlations remain unresolved.
We introduce a general representative-support framework for granular materials
based on three sequential requirements: statistical admissibility of the
sampled domain, persistent convergence of the apparent field mean and the
low-wavenumber covariance spectrum, and validation against independent
effective properties. The concept is tested on seven large-area
backscattered-electron images of Arabian dune sands and mineral-resolved
QEMSCAN maps. The structure-only criterion identifies a representative
elementary area of \(1536\) pixels, corresponding to approximately
\(2.01~\mathrm{mm}\). Without property-specific fitting or recalibration, the
same physical scale transferred to the QEMSCAN grid
(\(\approx 2.04~\mathrm{mm}\)) coincides with the onset of weak size dependence
in apparent thermal conductivity, elastic stiffness, and directional Young's
moduli. Thus, the representative scale inferred from long-wavelength granular
organization independently predicts the convergence of distinct constitutive
responses. The results show that representative support should be assigned
only after both statistical compatibility and long-range structural convergence
have been established. Although demonstrated here on two-dimensional dune-sand
sections, the criterion is formulated for granular materials generally and has
a direct mathematical extension to three-dimensional voxelized volumes.
\end{abstract}


\end{frontmatter}


\section{Introduction}
\label{sec:introduction}

Representative elementary area (REA) and representative elementary volume
(REV) selection is a fundamental scale-bridging problem in granular materials.
Sands, soils, fragmented rocks, powders, and other heterogeneous particulate
media contain structural organization over multiple length scales through
particle packing, phase distribution, pore geometry, contacts, compositional
heterogeneity, cracks, and cementation. Nevertheless, continuum-scale thermal,
hydraulic, mechanical, and transport properties are commonly inferred from
finite images or reconstructed volumes
\cite{Bear1972,Torquato2002,Blunt2013,AnovitzCole2015,Fu2021Tortuosity}.
The reliability of those effective properties therefore depends directly on
whether the observed support is representative of the material population to
which the homogenized response is assigned.

The usual operational procedure is to increase the observation window until a
selected scalar quantity---for example porosity, phase fraction, permeability,
conductivity, or stiffness---appears to reach a plateau
\cite{AlRaoushWillson2005,AlRaoushPapadopoulos2010,
CostanzaRobinson2011,Blunt2013,Andra2013a,Andra2013b}. This criterion is
convenient but incomplete. Two finite domains may exhibit nearly identical
porosity, phase fraction, or mean image intensity while retaining different
particle clustering, pore connectivity, anisotropy, and long-wavelength
organization \cite{Torquato2002,JiaoTorquato2007,JiaoTorquato2008}. These
differences can generate distinct heat-transfer pathways, load-transfer
networks, tortuosities, and capillary responses
\cite{Fu2021Tortuosity,TantardiniAlonsoMarroquin2026Capillarity}. Stabilization
of a first-order or property-specific scalar therefore does not, by itself,
establish that the underlying granular structure has become representative.

A second difficulty is spatial non-stationarity. Natural and engineered
granular materials need not form a single statistical population over the
entire imaged field. Packing density, particle size, mineralogy or composition,
pore fraction, segregation, cementation, and preparation or imaging response
may vary spatially. If statistically distinct domains are pooled, larger
windows can generate an apparently stable average simply by mixing different
populations. Under these conditions, increasing the support does not create
representativeness; it can instead conceal incompatibility. A representative
support should therefore be sought only inside a statistically admissible
material domain.

These observations motivate a different hierarchy for representative-support
selection. Statistical compatibility should be established first; structural
convergence should then be assessed using information that remains sensitive
to long-range organization; only afterwards should the predicted support be
tested against effective physical properties. Recent work has connected
representative sizing with covariance, spectral density, particle-scale
statistics, phase contrast, and macroscopic screening lengths
\cite{TantardiniAlonsoMarroquin2026Capillarity,
AlonsoMarroquinTantardini2026CylindricalREV}. In particular, the
slowest-converging spatial correlations provide a natural measure of whether
extended particle clusters, pore-rich domains, and slowly varying
compositional organization are fully represented.

Here we formulate representative support in granular materials through three
sequential conditions. First, \emph{statistical admissibility} requires the
candidate windows to belong to a common material population. Second,
\emph{long-wavelength structural convergence} requires both the apparent field
mean and the low-wavenumber covariance spectrum to remain converged throughout
the larger-window tail. Third, \emph{independent physical validation} tests
whether constitutive responses that were not used to select the structural
scale become weakly dependent on support size at the predicted length. The
resulting scale is therefore selected from microstructural statistics rather
than calibrated independently for each effective property.

We test this general concept on Arabian dune sands from the Dammam--Jafurah
sector, which provide a natural granular system with spatial variations in
packing, pore space, and mineralogical composition. Seven approximately
\(20~\mathrm{mm}\times20~\mathrm{mm}\) BSE images are used to identify
statistically compatible domains and determine the minimum support for which
the mean field and low-wavenumber covariance structure are persistently
converged. The predicted structural length is then transferred, without fitting
or property-specific recalibration, to independent QEMSCAN-derived thermal and
elastic property fields.

The dune-sand validation yields a structural REA of approximately
\(2.01~\mathrm{mm}\). More importantly, this scale independently predicts the
onset of weak size dependence in apparent thermal conductivity, elastic
stiffness, and directional Young's moduli. The central result is therefore not
a material-specific REA value, but a general criterion: representative support
in granular materials emerges only when the sampled domain is statistically
admissible and its long-wavelength structure has converged. The dune sands
serve as the experimental and computational test of that broader principle.

\section{Materials and methods}
\label{sec:methods}
The theory is operationalized through three sequential tests. Statistical
admissibility is established by detecting domains with compatible local
first- and second-order statistics. Structural representativeness is then
determined from the persistent convergence of the apparent field mean and the
low-wavenumber covariance spectrum. Finally, physical representativeness is
tested by transferring the structurally predicted length to independent
QEMSCAN-derived property fields and evaluating finite-window thermal and elastic
homogenization. The criterion-defining parameters and computational settings
are specified below together with the corresponding stages of the workflow.

\subsection{Dune-sand samples and preparation}
\label{subsec:samples_sem_mapping_1}

Dune-sand samples were collected in the Eastern Province of Saudi Arabia near
King Fahd International Airport, Dammam, within the northern sector of the
Jafurah Sand Sea. The Jafurah is a major aeolian system extending along the
Arabian Gulf margin, in which sediment transport, deflation, and deposition
vary systematically along the regional wind-transport pathway
\cite{FrybergerEtAl1984}. Petrographic, heavy-mineral, and detrital-zircon
studies have shown that Jafurah sands form part of a compositionally
heterogeneous and extensively recycled Arabian sediment-routing system, with
contributions from interior Arabian sources and, particularly toward the Gulf
margin, additional feldspar-, carbonate-, and lithic-bearing components
\cite{GarzantiEtAl2013ArabianSand,GarzantiEtAl2017Transcontinental}.
Regional surveys of Saudi dune fields similarly document predominantly
quartz-rich sands accompanied by variable feldspar, carbonate, mica, and
accessory-mineral fractions
\cite{BenaafiAbdullatif2015}.

The samples analysed here were treated as material from a common regional
deposit, but no assumption of global statistical stationarity was imposed on
the resulting polished sections. Statistically compatible domains were instead
identified directly from the image fields before representative-volume
analysis. Grain morphology and contact characteristics of related dune-sand
particles have been reported previously \cite{alonso2025contact}.

The sand was treated with \(10\%\) hydrochloric acid to remove carbonate
coatings and acid-soluble surface cement, rinsed with deionized water, and dried
before embedding. The cleaned grains were vacuum-impregnated in resin,
mechanically polished to expose a planar section, and carbon-coated before
electron-microscopy measurements.

\subsection{BSE imaging and QEMSCAN mineralogical mapping}
\label{subsec:samples_sem_mapping_2}

Seven large BSE images were acquired using a FEI Map 2.1 software operated at an accelerating voltage of
15 kV, a beam current of 10 nA, and a working
distance of 13 mm. Each full image covered approximately
\(20~\mathrm{mm}\times20~\mathrm{mm}\) and contained
\(15250\times15250\) pixels, giving an in-plane pixel size of approximately
\(1.31~\mu\mathrm{m}\).

Automated mineralogical maps were acquired from
\(10~\mathrm{mm}\times10~\mathrm{mm}\) regions using a QEMSCAN 650F instrument. QEMSCAN provides phase-resolved mineral maps, whereas BSE imaging
provides the larger field of view required for structural convergence analysis
\cite{ayling2012qemscan,Fu2023AutomatedMineralogy}.

BSE intensity is governed primarily by average atomic number. Brighter regions
therefore generally correspond to phases containing heavier elements, whereas
darker regions include low-\(Z\) silicates, pores, cracks, resin-filled spaces,
and topographic depressions. The BSE images were analysed as gray-level fields
that encode mineralogical and textural heterogeneity; they were not interpreted
as direct quantitative chemical-composition maps.

\subsection{Conversion of QEMSCAN maps into property fields}
\label{subsec:chemical_map_to_property_field}

The QEMSCAN outputs were represented as color-coded mineral maps on a regular
pixel grid. Each pixel was assigned to the closest reference color in the
mineral legend using a prescribed RGB-distance tolerance. This tolerance
accounts for antialiasing and minor rendering or compression variations. Pixels
that did not satisfy the matching tolerance were assigned to an unknown class.

The resulting mineral-index map was converted into spatially resolved thermal
and elastic property fields using Table~\ref{tab:mineral_database}. The same
mineral classification was used to construct
\[
\kappa(\mathbf{x}),\qquad
E(\mathbf{x}),\qquad
\nu(\mathbf{x}),
\]
where \(\kappa\) is the local thermal conductivity, \(E\) is Young's modulus,
and \(\nu\) is Poisson's ratio. The elastic parameters were subsequently used
to construct the local isotropic stiffness tensor
\(\mathbb{C}(\mathbf{x})\) under the plane-strain assumption.

The assigned mineral properties are representative microscopic values compiled
from published data for the thermal conductivity and elastic properties of
rock-forming minerals \cite{ClauserHuenges1995,Bass1995}. They were not fitted
to the homogenized responses and were kept fixed for every image, window size,
and loading condition.

Pore and background pixels were assigned air properties for thermal conduction
and a small nonzero stiffness for numerical stability. Pixels that could not be
matched to a mineral class within the prescribed color tolerance were assigned
to the unknown class and given the fallback values listed in
Table~\ref{tab:mineral_database}. All assigned properties were kept fixed
throughout the finite-window analysis.

This conversion separates mineral identification from constitutive assignment.
The QEMSCAN map determines the spatial phase distribution, whereas the
independent property database determines the coefficients used in thermal and
elastic homogenization. The full image-conversion workflow and RGB legend are
shown in Supplementary Fig.~S1 and Table~S2.

\begin{table}[t]
\centering
\caption{Constitutive properties assigned to the mineral classes used to construct the QEMSCAN-derived property fields. Thermal conductivity is reported in \(\mathrm{W\,m^{-1}\,K^{-1}}\), Young's modulus in GPa, and Poisson's ratio is dimensionless. The complete RGB segmentation table and image-conversion workflow are provided in Supplementary Note~S2.}
\label{tab:mineral_database}
\small
\begin{tabular}{lrrr}
\hline
Mineral & \(\kappa\) & \(E\) & \(\nu\) \\
\hline
Quartz           & 6.5   & 94    & 0.08 \\
K-Feldspar       & 2.3   & 70    & 0.25 \\
Plagioclase      & 2.0   & 76    & 0.26 \\
Mica             & 1.0   & 60    & 0.25 \\
Calcite          & 3.6   & 72    & 0.31 \\
Dolomite         & 5.0   & 94    & 0.28 \\
Clay minerals    & 1.5   & 20    & 0.30 \\
Gypsum/Anhydrite & 2.5   & 45    & 0.30 \\
Heavy minerals   & 5.0   & 100   & 0.25 \\
Pores/Background & 0.026 & 0.001 & 0.20 \\
Others/unknown   & 0.0   & 0.001 & 0.20 \\
\hline
\end{tabular}
\end{table}

\subsection{Statistical-admissibility and stationary-domain detection}
\label{subsec:stationary_domain_detection}

The present analysis concerns two-dimensional polished sections. The reported
conductivity, stiffness, and directional Young moduli are therefore in-plane
apparent properties and are not interpreted as bulk three-dimensional material
properties.
Statistical admissibility requires candidate REA windows to sample a common
material population. We therefore identify approximately stationary domains
before examining convergence with observation size. This operational test
selects regions having compatible local means and fluctuation amplitudes, but
does not by itself establish weak stationarity of the complete image field.
The convergence of the long-wavelength second-order structure is assessed
separately through the covariance-spectral criterion introduced below.

Let \(Z_i(\mathbf{x})\) be the BSE gray-level field in the \(i\)-th independent
image sample, with \(i=1,\ldots,N_s\). Here \(N_s=7\). Each image is scanned
with square stationarity windows of side length \(R_{\rm stat}\). These
windows are used only to detect locally compatible image regions; they are not
candidate REA windows.

For each stationarity window \(W_p^{(i)}\), we compute the local mean and local
standard deviation,
\begin{align}
\mu_p^{(i)}
&=
\frac{1}{|W_p^{(i)}|}
\int_{W_p^{(i)}} Z_i(\mathbf{x})\,d\mathbf{x},
\nonumber\\
\sigma_p^{(i)}
&=
\left[
\frac{1}{|W_p^{(i)}|}
\int_{W_p^{(i)}}
\left(
Z_i(\mathbf{x})-\mu_p^{(i)}
\right)^2
d\mathbf{x}
\right]^{1/2}.
\label{eq:local_stationarity_mean_std}
\end{align}
The robust image-level references are the medians of these local quantities,
\begin{equation}
\mu_{\rm med}^{(i)}
=
\operatorname{median}_{p}\mu_p^{(i)},
\qquad
\sigma_{\rm med}^{(i)}
=
\operatorname{median}_{p}\sigma_p^{(i)} .
\label{eq:stationarity_reference_medians}
\end{equation}
A stationarity window is accepted when both its mean and standard deviation are
close to the image-level references,
\begin{align}
\frac{
\left|
\mu_p^{(i)}-\mu_{\rm med}^{(i)}
\right|
}{
\left|
\mu_{\rm med}^{(i)}
\right|+\epsilon
}
&\le
\tau_{\mu},
\nonumber\\
\frac{
\left|
\sigma_p^{(i)}-\sigma_{\rm med}^{(i)}
\right|
}{
\left|
\sigma_{\rm med}^{(i)}
\right|+\epsilon
}
&\le
\tau_{\sigma}.
\label{eq:stationary_window_acceptance}
\end{align}
Here \(\tau_\mu\) and \(\tau_\sigma\) are the prescribed compatibility
tolerances, while \(\epsilon>0\) is a numerical regularizer that prevents
division by zero. Equations~\eqref{eq:local_stationarity_mean_std} and
\eqref{eq:stationary_window_acceptance} test only the local compatibility of
the first two marginal moments and are used to construct the statistically
admissible-domain mask.

The accepted stationarity windows were converted into a binary pixel mask
\(S_i(\mathbf{x})\in\{0,1\}\), where \(S_i(\mathbf{x})=1\) identifies pixels
belonging to locally compatible regions.

The resulting stationary domain was defined as
\begin{equation}
\Omega_{\rm stat}^{(i)}
=
\left\{
\mathbf{x}\;:\;S_i(\mathbf{x})=1
\right\}.
\label{eq:stationary_domain_def}
\end{equation}
The structural REA was therefore determined within the detected stationary
domains rather than assigned globally to the complete BSE image.

The admissibility procedure is designed to localize statistically compatible
regions in a two-dimensional material field. Conventional unit-root tests such
as ADF, Phillips--Perron, and KPSS act on ordered one-dimensional sequences and
do not directly provide a traversal-independent connected spatial domain.
Wiener--Khinchin consistency addresses second-order spectral consistency within
a prescribed domain but does not itself localize compatible subdomains. A
detailed comparison of these diagnostics with the present spatial criterion is
provided in Supplementary Note~S1 and Table~S1.

\subsection{Structural REA criterion from persistent mean--spectral convergence}
\label{sec:rev_multiphysics}

Structural representativeness requires a candidate support to reproduce both the
average field level and the longest spatial correlations of the stationary
material domain. Candidate square supports were therefore tested using a joint
criterion based on the apparent gray-level mean and the low-wavenumber
covariance spectrum.
The tested square side lengths form the ordered set
\begin{equation}
\mathcal{L}
=
\left\{
L_1,L_2,\ldots,L_M
\right\},
\qquad
L_1<L_2<\cdots<L_M .
\label{eq:L_sequence_rev}
\end{equation}

For the full-resolution BSE analysis, the tested support sequence was
\begin{equation}
\mathcal L=\{L_n\}_{n=1}^{15},
\qquad
L_n=
\begin{cases}
128(n+1), & 1\le n\le 7,\\
256(n-3), & 8\le n\le 11,\\
512(n-7), & 12\le n\le 15,
\end{cases}
\quad \mathrm{pixels}.
\label{eq:BSE_tested_support_sequence}
\end{equation}
The largest tested support,
\begin{equation}
L_{\rm ref}=4096~\mathrm{pixels},
\label{eq:BSE_reference_support}
\end{equation}
was used as the finite observational reference.

The physical side length associated with \(L_m\) is
\begin{equation}
\ell_m=L_m\Delta x ,
\label{eq:pixel_physical_window_size}
\end{equation}
where \(\Delta x\) is the BSE pixel size. The reference support is finite and
is not interpreted as an infinite-volume limit.

For each \(L\in\mathcal L\), a candidate window
\(\Omega_{L,p}^{(i)}\) is retained only if it lies almost entirely inside the
stationary domain,
\begin{equation}
\frac{
\left|
\Omega_{L,p}^{(i)}
\cap
\Omega_{\rm stat}^{(i)}
\right|
}{
\left|
\Omega_{L,p}^{(i)}
\right|
}
\ge
f_{\rm stat}.
\label{eq:min_mask_fraction_rule}
\end{equation}

For each support size, candidate windows were generated deterministically on a
regular \(7\times7\) spatial grid over each image. The centred window was also
included explicitly when it was not already present on this grid. Candidate
windows were retained only when they satisfied
Eq.~\eqref{eq:min_mask_fraction_rule}. Consequently, the number of accepted
windows could vary with support size and image, with at most 50 distinct
candidate positions before stationary-domain filtering. The same deterministic
sampling protocol was used for all seven images.

This condition ensures that the finite-window sequence is built from one
statistically compatible population.
A support size was retained in the common REA scan only when at least one
window satisfying Eq.~\eqref{eq:min_mask_fraction_rule} was available in each
of the seven images.

The first part of the REA criterion tests the apparent level of the field. For
each accepted window, the apparent gray-level mean is
\begin{equation}
\langle Z\rangle_{L,p}^{(i)}
=
\frac{1}{|\Omega_{L,p}^{(i)}|}
\int_{\Omega_{L,p}^{(i)}} Z_i(\mathbf{x})\,d\mathbf{x}.
\label{eq:apparent_mean_Z_window}
\end{equation}
The corresponding image-level and ensemble-level means are
\begin{equation}
\langle Z\rangle_L^{(i)}
=
\frac{1}{N_L^{(i)}}
\sum_{p=1}^{N_L^{(i)}}
\langle Z\rangle_{L,p}^{(i)},
\qquad
\overline Z_L
=
\frac{1}{N_s}
\sum_{i=1}^{N_s}
\langle Z\rangle_L^{(i)}.
\label{eq:ensemble_mean_Z}
\end{equation}
The apparent-mean residual is then
\begin{equation}
r_Z(L)
=
\frac{
\left|
\overline Z_L-\overline Z_{L_{\rm ref}}
\right|
}{
\left|
\overline Z_{L_{\rm ref}}
\right|+\epsilon
}.
\label{eq:raw_Z_residual}
\end{equation}
This quantity measures whether the mean field level at support \(L\) is still
biased relative to the largest valid stationary support.

The second part of the criterion tests spatial organization. 

For each image and support size, one admissible stationary window was selected
for the spectral diagnostic. Its local mean was removed, and the
two-dimensional discrete Fourier transform of the resulting field was
evaluated. The covariance spectrum was obtained from the squared Fourier
amplitude. 
The spectrum was radially averaged into \(J\) bins and
interpolated onto the common wavenumber grid
\[
0=q_1<q_2<\cdots<q_J\le k_{\max},
\qquad
k_{\max}=\frac{\pi}{\Delta x}.
\]
Only the band \(\mathcal K_{\rm low}\) was used for REA selection because it
contains the longest spatial fluctuations resolved by the image.

The selected window was the admissible stationary window whose centre was
closest to the image centre. The same deterministic rule was applied at every
support size, including the finite reference support \(L_{\rm ref}\). The
spectral residual was evaluated independently for each image and subsequently
averaged over the seven-image ensemble.

The common radial grid was uniformly spaced and trapezoidal quadrature weights
\(w_j^{\rm trap}\) were used to approximate the spectral integrals. To give
greater weight to the longest resolved spatial fluctuations, the quadrature
weights were modified according to
\begin{equation}
w_j
=
\frac{w_j^{\rm trap}}
{q_j+\epsilon_k},
\qquad
\epsilon_k=q_2,
\label{eq:lowk_spectral_weights}
\end{equation}
where \(q_2\) is the smallest positive wavenumber on the common grid. Because
the same weights enter the numerator and denominator of the normalized
residual, no additional normalization of \(w_j\) is required.

To quantify convergence of the low-wavenumber covariance spectrum, we define
the weighted discrete norm
\begin{equation}
\left\|f\right\|_{w,\mathcal K_{\rm low}}
=
\left[
\sum_{q_j\in\mathcal K_{\rm low}}
w_j f^2(q_j)
\right]^{1/2}.
\label{eq:weighted_lowk_norm}
\end{equation}
A scale-dependent numerical floor is introduced as
\begin{equation}
\epsilon_{\widehat C}
=
\epsilon_{\rm rel}
\max_{r=1,\ldots,N_s}
\left\|
\widehat C_{L_{\rm ref}}^{(r)}
\right\|_{w,\mathcal K_{\rm low}},
\qquad
\epsilon_{\rm rel}=10^{-15}.
\label{eq:spectral_regularization}
\end{equation}
The image-level low-\(k\) residual is then
\begin{equation}
\eta_{\widehat C}^{(i)}(L)
=
\frac{
\left\|
\widehat C_L^{(i)}
-
\widehat C_{L_{\rm ref}}^{(i)}
\right\|_{w,\mathcal K_{\rm low}}
}{
\max\left[
\left\|
\widehat C_{L_{\rm ref}}^{(i)}
\right\|_{w,\mathcal K_{\rm low}},
\epsilon_{\widehat C}
\right]
},
\qquad
L<L_{\rm ref}.
\label{eq:eta_C_lowk_image}
\end{equation}
The corresponding ensemble residual is
\begin{equation}
\eta_{\widehat C}(L)
=
\frac{1}{N_s}
\sum_{i=1}^{N_s}
\eta_{\widehat C}^{(i)}(L).
\label{eq:eta_C_lowk_discrete}
\end{equation}

This residual does not test whether an individual spectrum becomes flat as a
function of wavenumber. Instead, it tests whether the complete low-wavenumber
spectral content becomes invariant with respect to further increases in the
observation support. Two spectra may each appear approximately flat while
retaining different amplitudes or low-\(k\) shapes. The relative weighted norm
therefore provides a quantitative and reproducible measure of the visual
spectral collapse. The use of low-wavenumber spectral convergence for
representative-support sizing follows
Ref.~\cite{AlonsoMarroquinTantardini2026CylindricalREV}.

Because finite-window curves can contain isolated fluctuations, the
apparent-mean residual is regularized using a centred five-point running median
over the ordered support sequence. The finite reference point is excluded
because its residual is zero by construction. Near the ends of the
non-reference sequence, the median is evaluated using the available neighbouring
values. The resulting regularized sequence,
\(\widetilde r_Z(L_m)\), is summarized by the persistent tail envelope
\begin{equation}
\eta_Z^{\rm med-tail}(L_m)
=
\max_{L_j\in\mathcal{L},\,L_j\ge L_m,\,L_j<L_{\rm ref}}
\widetilde r_Z(L_j).
\label{eq:eta_Z_med_tail}
\end{equation}
The selected structural REA is the smallest tested support for which both the
apparent-mean tail envelope and the low-\(k\) spectral residual remain below
their tolerances over the non-reference tail:
\begin{equation}
\begin{aligned}
L_{\rm REA}
=
\min\bigg\{\,L_m\in\mathcal{L},\; L_m<L_{\rm ref}:\;&
\eta_Z^{\rm med-tail}(L_m)\le\tau_Z,\\
&\eta_{\widehat C}(L')\le\tau_{\widehat C}
\quad\forall\,L'\in\mathcal{L},\\
&L'\ge L_m,\; L'<L_{\rm ref}\bigg\}.
\end{aligned}
\label{eq:spectral_rev_rule}
\end{equation}

The reference \(L_{\rm ref}\) is the largest support for which the same
stationary-domain sampling rule can be applied across the image ensemble. It is
a finite observational reference and is not interpreted as an infinite-volume
limit. A candidate \(L_m\) was evaluated only when at least one half of the
non-reference support sequence remained above it. This restriction prevents
selection from an under-sampled tail immediately below \(L_{\rm ref}\).

No smoothing was applied to the low-wavenumber covariance residual used in the
REA-selection criterion.

The physical REA length is then
\begin{equation}
\ell_{\rm REA}=L_{\rm REA}\Delta x .
\label{eq:physical_rev_length}
\end{equation}

Additional ensemble-reproducibility diagnostics, which are evaluated only after the REA has been selected and do not enter the acceptance rule, are given in Supplementary Note~S3.

\subsection{Transfer of the structural scale to independent property maps}
\label{subsec:property_map_stationarity}

Physical validation was performed on QEMSCAN-derived property fields that are
independent of the BSE gray-level statistics used to select the structural REA.
The BSE-derived physical length was transferred directly to the property-map
grid without fitting or property-specific recalibration. Stationary-domain
filtering was repeated on the property fields to ensure that the finite-element
windows sampled statistically compatible regions on the QEMSCAN grid.

For each image sample, the conductivity, Young-modulus, and Poisson-ratio fields
were used as numerical property fields rather than rescaled visualization images.
They were robustly standardized and combined into a common property-stationarity
descriptor. The same local mean--standard-deviation admissibility test used for
the BSE field was then applied on the QEMSCAN grid. The complete robust
standardization, including the median-absolute-deviation scaling and fallback
rules, is reported in Supplementary Note~S3.

The BSE-derived structural length was converted to the QEMSCAN grid according
to
\begin{equation}
L_{\rm REA}^{\rm prop,raw}
=
\frac{\ell_{\rm REA}}{\Delta x_{\rm prop}},
\label{eq:Lrev_property_grid_general}
\end{equation}
where \(\Delta x_{\rm prop}\) is the QEMSCAN pixel size. For the present data,
\[
L_{\rm REA}^{\rm prop,raw}
=
\frac{2.012~\mathrm{mm}}
{0.01~\mathrm{mm\,pixel^{-1}}}
=
201.2~\mathrm{pixels}.
\]
Because the property-level finite-element scan was performed on a discrete set
of support sizes, the directly converted value was mapped to the nearest tested
support, giving \(L_{\rm REA}^{\rm prop}=204\) pixels.

Homogenization was restricted to windows satisfying the property-domain mask
coverage criterion. At most four accepted windows were analysed at each support,
compared with up to 50 candidate windows in the BSE structural analysis. The
thermal and elastic curves therefore provide an independent physical validation
of the predicted structural scale rather than a second high-statistics estimate
of the REA.

\subsection{Finite-window periodic homogenization}
\label{subsec:periodic_homogenization}

Property-level validation was performed by solving periodic cell problems on
square windows of side length \(L\). Each accepted window defines a periodic
cell \(\Omega(L)=[0,L)^2\), with one bilinear quadrilateral finite element per
pixel and voxelwise constant material properties. Apparent tensors were obtained
from volume-averaged fluxes or stresses. Full discretization and solver details
are provided in Supplementary Note~S3.

For thermal conduction, the scalar potential is decomposed as
\begin{equation}
u(\mathbf{x})=\mathbf E\cdot\mathbf{x}+\widetilde u(\mathbf{x}),
\qquad
\langle\widetilde u\rangle_{\Omega(L)}=0,
\label{eq:scalar_affine_periodic_methods}
\end{equation}
where \(\widetilde u\) is periodic. The corrector problem is
\begin{equation}
-\nabla\cdot\left[\kappa(\mathbf{x})
\left(\mathbf E+\nabla\widetilde u\right)\right]=0
\qquad \text{in }\Omega(L),
\label{eq:scalar_cell_problem_methods}
\end{equation}
and the apparent-conductivity tensor follows from the volume-averaged flux,
\begin{equation}
\boldsymbol{\kappa}_{\rm app}^{(:,j)}(L)=
\left\langle\kappa(\mathbf{x})
\left[\widehat{\mathbf e}_j+\nabla\widetilde u^{(j)}\right]
\right\rangle_{\Omega(L)}.
\label{eq:kappa_app_methods}
\end{equation}
The reported directional components are
\(\kappa_x^{\rm app}=\kappa_{11}^{\rm app}\) and
\(\kappa_y^{\rm app}=\kappa_{22}^{\rm app}\).

For elasticity, local isotropic properties \(E(\mathbf{x})\) and
\(\nu(\mathbf{x})\) define the plane-strain stiffness tensor
\(\mathbb C(\mathbf{x})\). The displacement is decomposed into an imposed
macroscopic strain and a periodic fluctuation,
\begin{equation}
\mathbf u(\mathbf{x})=
\overline{\boldsymbol\varepsilon}\,\mathbf{x}
+\widetilde{\mathbf u}(\mathbf{x}),
\qquad
\langle\widetilde{\mathbf u}\rangle_{\Omega(L)}=\mathbf 0,
\label{eq:elastic_affine_periodic_methods}
\end{equation}
with corrector problem
\begin{equation}
-\nabla\cdot\left\{\mathbb C(\mathbf{x}):
\left[\overline{\boldsymbol\varepsilon}
+\boldsymbol\varepsilon(\widetilde{\mathbf u})\right]\right\}=\mathbf 0
\qquad \text{in }\Omega(L).
\label{eq:elastic_cell_problem_methods}
\end{equation}
The apparent stiffness tensor is defined by
\begin{equation}
\mathbb C_{\rm app}(L):\overline{\boldsymbol\varepsilon}
=
\left\langle
\mathbb C(\mathbf{x}):
\left[\overline{\boldsymbol\varepsilon}
+\boldsymbol\varepsilon(\widetilde{\mathbf u})\right]
\right\rangle_{\Omega(L)}.
\label{eq:Capp_methods}
\end{equation}
In two-dimensional Voigt notation, the apparent compliance matrix
\(\mathbf S_{\rm app}=\mathbf C_{\rm app}^{-1}\) gives
\begin{equation}
E_x^{\rm app}(L)=\frac{1}{S_{11}^{\rm app}(L)},
\qquad
E_y^{\rm app}(L)=\frac{1}{S_{22}^{\rm app}(L)}.
\label{eq:Eapp_methods}
\end{equation}

\subsection{Statistical analysis}
\label{subsec:statistical_analysis}

The structural analysis used seven independent BSE images. Image-level
statistics were computed independently for each image and subsequently averaged
over the ensemble, ensuring that each image contributed equally regardless of
the number of admissible stationary windows. The structural REA was determined
solely from the persistent mean--spectral criterion in
Eq.~\eqref{eq:spectral_rev_rule}. Ensemble reproducibility and property-level
homogenization were not included as acceptance constraints in that criterion
and were used as validation diagnostics.

\subsection{Computational implementation}
\label{subsec:computational_implementation}

Stationary-domain analysis, finite-window sampling, covariance-spectrum
evaluation, and REA selection were implemented in Python using in-house
software. Periodic thermal and elastic homogenization was performed with
in-house finite-element solvers applied to the QEMSCAN-derived property fields.
The source code, parameter files, and processing scripts required to reproduce
the stationary masks, covariance spectra, REA curves, and homogenized-property
calculations are publicly archived as specified in the Data and code availability
statement.

The criterion-defining numerical parameters are collected in
Table~\ref{tab:rev_numerical_parameters}. These values specify the statistical
admissibility and structural REA decision rule used in the main analysis. More
detailed implementation choices, including the complete support sequence,
radial-grid discretization, sampling counts, and auxiliary numerical settings,
are reported in Supplementary Table~S3.

\begin{table}[!ht]
\centering
\caption{Criterion-defining numerical parameters used for statistical admissibility and structural REA selection. The full implementation parameter set is reported in Supplementary Table~S3.}
\label{tab:rev_numerical_parameters}
\small
\begin{tabular}{p{0.28\linewidth}p{0.18\linewidth}p{0.44\linewidth}}
\hline
Parameter & Value & Role \\
\hline
\(\tau_\mu\) & \(0.02\) & Local-mean compatibility tolerance \\
\(\tau_\sigma\) & \(0.06\) & Local-standard-deviation compatibility tolerance \\
\(\tau_Z\) & \(0.025\) & Apparent-mean residual tolerance \\
\(\tau_{\widehat C}\) & \(0.30\) & Low-wavenumber covariance-spectrum tolerance \\
\(f_{\rm stat}\) & \(0.98\) & Minimum admissible-mask fraction in retained windows \\
\(R_{\rm stat}\) & \(2048\) px & Stationarity-window side length \\
\(\mathcal K_{\rm low}\) & \(k\le0.25\,k_{\max}\) & Spectral band used for structural convergence \\
\(L_{\rm ref}\) & \(4096\) px & Largest finite observational reference support \\
\hline
\end{tabular}
\end{table}

\section{Results and discussion}
\label{sec:results_discussion}

The results are presented in the same logical order as the proposed
representative-support criterion. We first establish the statistically
admissible granular domains in which representative sizing is meaningful. We
then determine the minimum support that persistently reproduces the
long-wavelength structural content of that population. Finally, we test whether
the structurally predicted scale, obtained without constitutive information,
also marks the convergence of independent thermal and elastic responses in the
dune-sand validation system.

\subsection{Statistical admissibility excludes incompatible granular domains}
\label{subsec:stationary_domains}

Before determining the size of a representative elementary area or volume, it
is necessary to establish whether the sampled field contains a statistically
compatible material population for which such a representative support can be
defined. This requirement is independent of spatial dimensionality and applies
to both two-dimensional images and three-dimensional voxelized volumes.
Conventional finite-window analyses often assume implicitly that all sampled
subdomains belong to the same statistical population. In heterogeneous
materials, however, spatial variations in composition, phase distribution,
porosity, particle size, morphology, packing, sample preparation, or imaging
response may violate this assumption. Under such conditions, an apparent
large-support plateau may result from averaging statistically distinct domains
rather than from genuine structural representativeness.

The framework therefore identifies statistically compatible regions before
evaluating structural convergence. Local moving-window statistics are compared
with robust field-level reference values to construct a spatial admissibility
mask, as defined in
Eqs.~\eqref{eq:local_stationarity_mean_std}--\eqref{eq:stationary_window_acceptance}.
This test provides an operational measure of local compatibility in the first
two marginal moments. It is not, by itself, a sufficient proof of weak
stationarity; convergence of the spatial covariance structure is assessed
independently in the subsequent analysis.

The concept is formulated generally for spatially resolved material fields,
including three-dimensional volumetric data. Its numerical validation in the
present work is performed on two-dimensional BSE and QEMSCAN images, while the
corresponding three-dimensional formulation is given in
Supplementary Note~S4.

Figure~\ref{fig:stationarity_workflow} illustrates the two-dimensional
stationarity analysis for a representative BSE image. Although the original
gray-level field appears broadly homogeneous by visual inspection
[Fig.~\ref{fig:stationarity_workflow}(a)], the local-mean map reveals gradual
spatial variations that are not readily apparent in the raw image
[Fig.~\ref{fig:stationarity_workflow}(c)]. These variations show that some
regions do not share the same average imaging response and therefore should not
be treated as realizations of a single statistical population.

The normalized stationarity residual provides complementary information
[Fig.~\ref{fig:stationarity_workflow}(d)]. It identifies regions in which the
local mean, the local fluctuation amplitude, or both depart from the dominant
image-level statistics. Small residuals indicate compatibility with the
prescribed stationarity tolerances, whereas larger values identify statistically
incompatible regions. The residual can therefore distinguish domains having
similar average intensity but different levels of local heterogeneity.

Combining the local-mean and local-fluctuation criteria yields the
stationary-domain mask shown in Fig.~\ref{fig:stationarity_workflow}(b).
Regions satisfying both criteria are retained as statistically admissible,
whereas incompatible regions are excluded from the subsequent finite-window
analysis. Candidate REA windows are required to satisfy the prescribed
stationary-domain coverage threshold. The resulting convergence curves
therefore characterize a single statistically compatible material population
rather than an average over spatially distinct domains.

This two-dimensional result verifies the first stage of the framework:
statistical admissibility must be established before a representative support
is sought. Increasing the observation size cannot, by itself, make an
inhomogeneous field representative. When sampled subdomains belong to different
statistical populations, larger supports merely average their differences and
may generate a misleading apparent plateau. The REA determined below is
therefore defined within the detected stationary domains rather than as a
global property of the complete polished section. The same principle applies
in three dimensions, where a representative elementary volume can be assigned
only within statistically compatible subvolumes.

\begin{figure}[H]
\centering
\includegraphics[width=\linewidth]{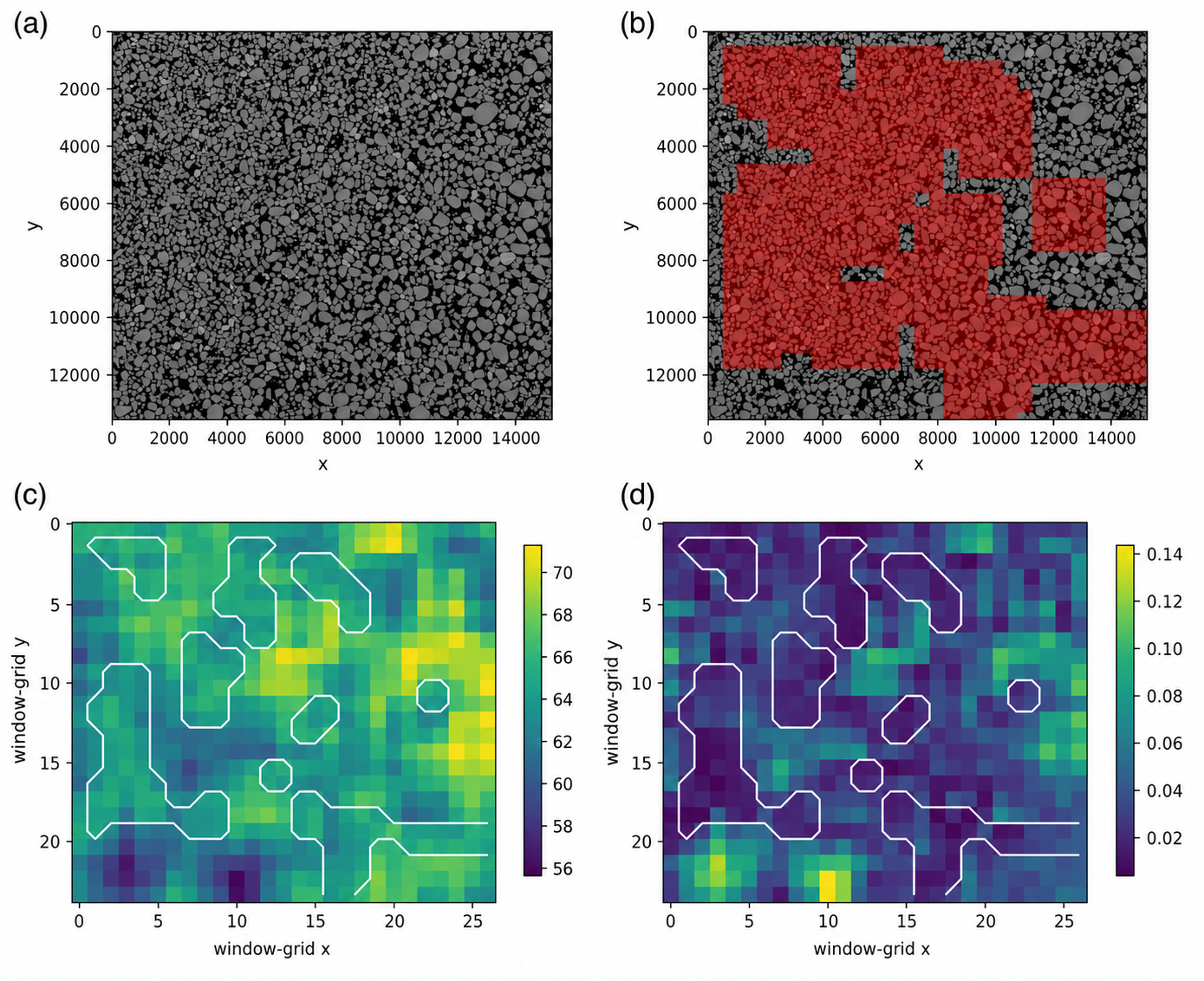}
\caption{
Identification of statistically stationary domains in a representative
backscattered-electron (BSE) image.
The figure tests the statistical-admissibility condition of the proposed REA
theory.
Before representative-volume sizing, the image is
tested for spatial stationarity to determine whether different regions belong
to the same statistical material population.
\textbf{(a)} Full-resolution BSE gray-level field.
\textbf{(b)} Stationary-domain mask obtained from the local stationarity test.
The highlighted regions satisfy the prescribed local statistical criteria and
define the domains in which REA analysis is performed.
\textbf{(c)} Spatial distribution of the local window mean,
\(\mu(\mathbf{x})\), computed over the stationarity windows. Slow spatial
variations indicate regions with different average gray-level statistics.
\textbf{(d)} Maximum normalized local residual used to assess stationarity.
Low residual values identify statistically compatible regions, whereas high
residual values indicate departures from stationarity. The white contour marks
the connected stationary domain retained for the subsequent REA analysis.
}
\label{fig:stationarity_workflow}
\end{figure}

\subsection{Long-wavelength structural convergence determines representative support}
\label{subsec:structural_rev}

Having restricted the analysis to statistically admissible domains, we next
determine the minimum observation support that reproduces the structural
statistics of the material population. Candidate windows are retained only
when they satisfy the stationary-domain coverage criterion in
Eq.~\eqref{eq:min_mask_fraction_rule}. Structural convergence is therefore
evaluated within a common statistical domain rather than across a mixture of
spatially incompatible regions.

\begin{figure}[H]
    \centering
    \includegraphics[width=\linewidth]{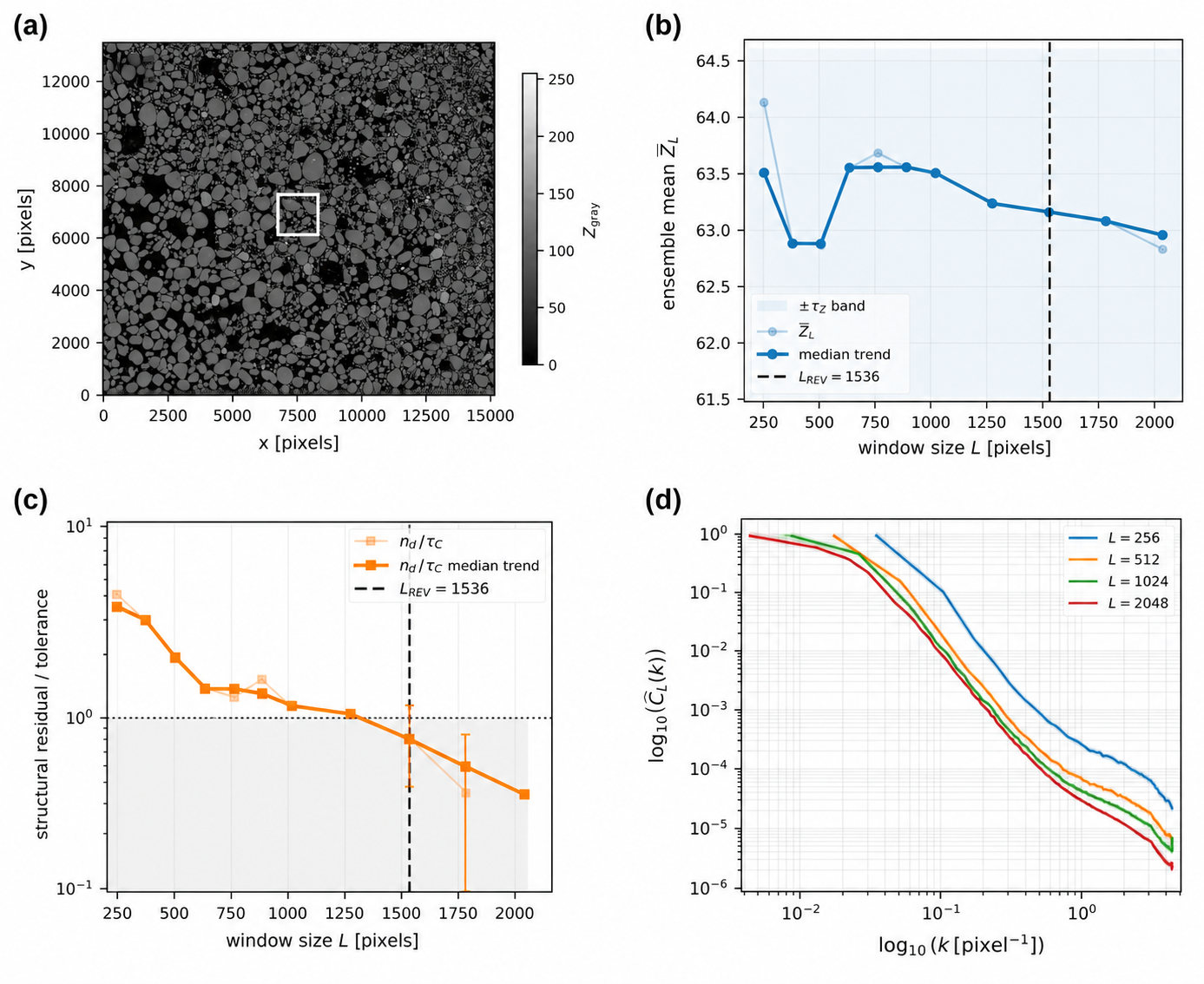}
    \caption{
    Structural-representativeness test based on persistent mean and
low-wavenumber covariance convergence.
    \textbf{(a)} Representative full-resolution BSE gray-level field
    \(Z_{\rm gray}(\mathbf{x})\). The white square marks the selected support,
    \(L_{\rm REA}=1536\) pixels. Candidate windows are evaluated only inside
    domains satisfying the stationary-mask coverage criterion.
    \textbf{(b)} Ensemble apparent gray-level mean \(\overline Z_L\) as a
    function of window size. Faint symbols show raw finite-window values, the
    solid curve is a median-trend guide, and the shaded region indicates the
    tolerance band used for the apparent-mean part of the REA rule.
    \textbf{(c)} Low-wavenumber covariance residual. Faint symbols show the raw
    normalized residual \(\eta_{\widehat C}/\tau_{\widehat C}\), while the solid
    curve is a median-trend guide. The horizontal dotted line marks the
    acceptance threshold and the vertical dashed line marks the selected REA.
    \textbf{(d)} Radially averaged covariance spectra \(\widehat C_L(k)\) for
    selected window sizes. The collapse of the low-wavenumber part of the
    spectrum indicates convergence of the long-wavelength structural content.
    }
    \label{fig:sand_rev_summary}
\end{figure}

Figure~\ref{fig:sand_rev_summary}(b) shows the evolution of the ensemble
gray-level mean, \(\overline{Z}_L\), with increasing window size. The apparent
mean approaches a stable large-window regime as local heterogeneities are
progressively averaged. Stabilization of the mean is necessary for
representativeness, but it is not sufficient. Windows with similar mean gray
levels may still contain different grain arrangements, pore networks, or
mineral assemblages. First-order statistics alone therefore cannot establish
that the spatial organization of the microstructure has converged.

We resolve this ambiguity by examining the covariance spectrum, which
quantifies the spatial organization of the mean-centred gray-level field. The
spectral residual and the persistent convergence criterion are defined in
Sec.~\ref{sec:rev_multiphysics}. The analysis focuses on the low-wavenumber
region because it represents the longest spatial fluctuations resolved within
the image. In the present dune-sand microstructure, these fluctuations include
extended grain clusters, pore-rich regions, and slowly varying mineralogical
organization. Such large-scale features generally require larger observation
supports to converge than local structures such as individual pores, grain
boundaries, or short-range intensity variations.

This scale separation is evident in
Fig.~\ref{fig:sand_rev_summary}(d). The high-wavenumber components become
similar at comparatively small window sizes, whereas substantial differences
persist at low wavenumber. The normalized low-wavenumber covariance residual
in Fig.~\ref{fig:sand_rev_summary}(c) summarizes this convergence
quantitatively. Small windows produce large residuals because they do not
capture the full long-range covariance structure of the stationary domain.
As the observation support increases, progressively longer spatial
correlations are incorporated and the residual approaches the prescribed
tolerance.

The structural REA is selected using the persistent mean--spectral criterion
introduced in Sec.~\ref{sec:rev_multiphysics}. The criterion does not identify
the first isolated threshold crossing. Instead, both the regularized
apparent-mean residual and the low-wavenumber covariance residual must remain
below their respective tolerances over the complete sequence of larger
non-reference windows. This persistence condition suppresses isolated
finite-sampling fluctuations and identifies the smallest support that
consistently reproduces both the average field level and the long-wavelength
spatial organization.

Application of this criterion to the seven independent BSE images yields
\[
L_{\rm REA}=1536\ \text{pixels},
\]
corresponding to a physical side length of approximately
\(2.01~\mathrm{mm}\). This value is not inferred from stabilization of the
apparent mean alone. It is the smallest tested support for which both the
first-order field statistic and the low-wavenumber covariance structure remain
converged throughout the larger-window tail.

The result provides the two-dimensional validation of the structural
representativeness criterion developed here. More generally, a representative
support is obtained only when it captures the longest spatial correlations
relevant to the material field. In three dimensions, the same principle
defines a structural REV by replacing planar windows with voxelized
subvolumes and the radial spectral average with the corresponding
three-dimensional shell average, as formulated in
Supplementary Note~S4.

Independent ensemble-reproducibility checks confirm that the selected two-dimensional support is not controlled by a single BSE image, one window location, or an isolated fluctuation; these auxiliary results are reported in Supplementary Note~S3 and Fig.~S2.

\subsection{Independent thermal and elastic convergence validates the predicted support}
\label{subsec:property_validation}

We next test whether the structurally defined support predicts the scale at
which independent constitutive responses become representative. The REA
obtained exclusively from the BSE gray-level statistics is transferred to the
QEMSCAN grid and evaluated against thermal and elastic homogenization, without
fitting or property-specific recalibration.

\begin{figure}[H]
    \centering
    \includegraphics[height=0.35\linewidth]{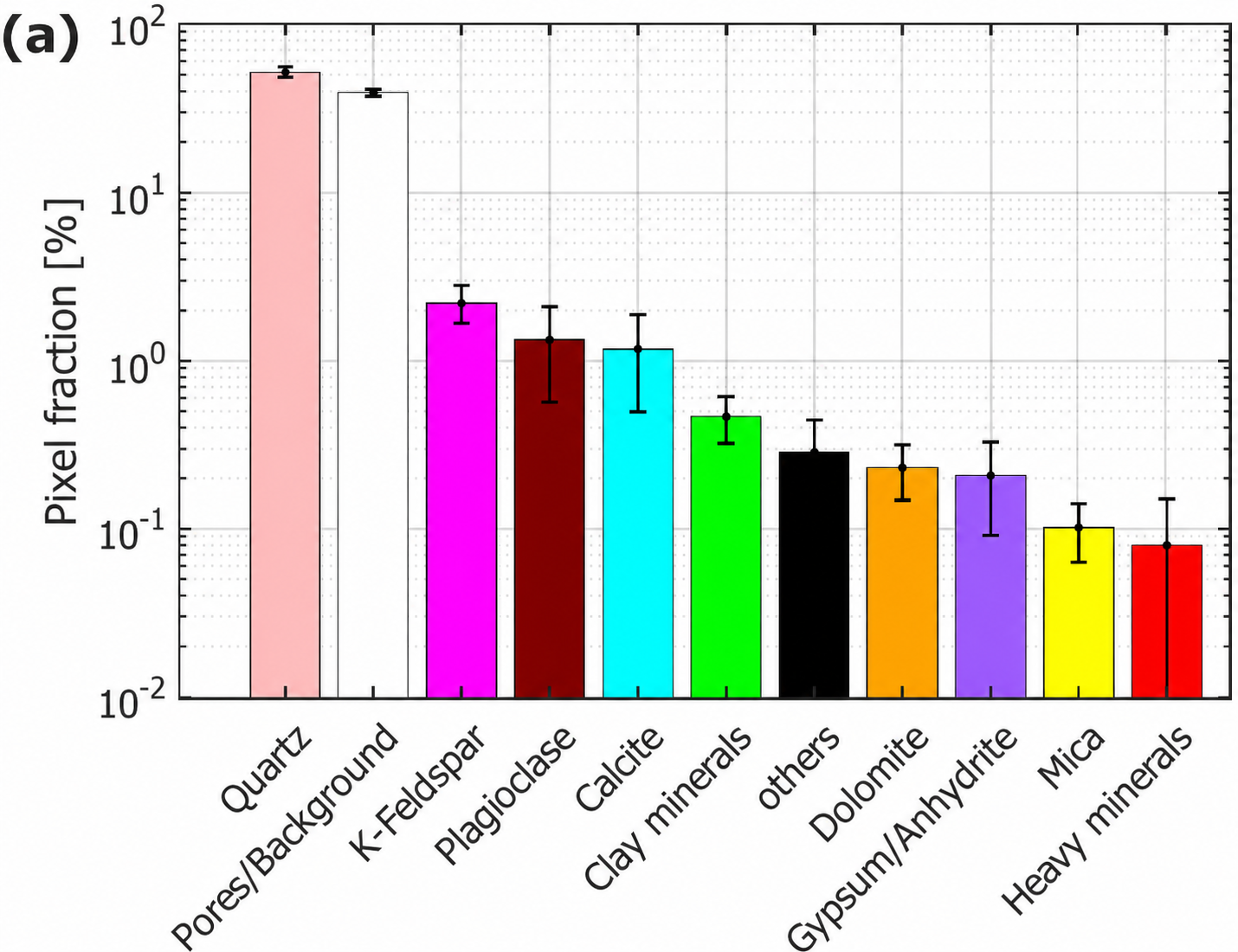}
    \includegraphics[width=0.48\linewidth]{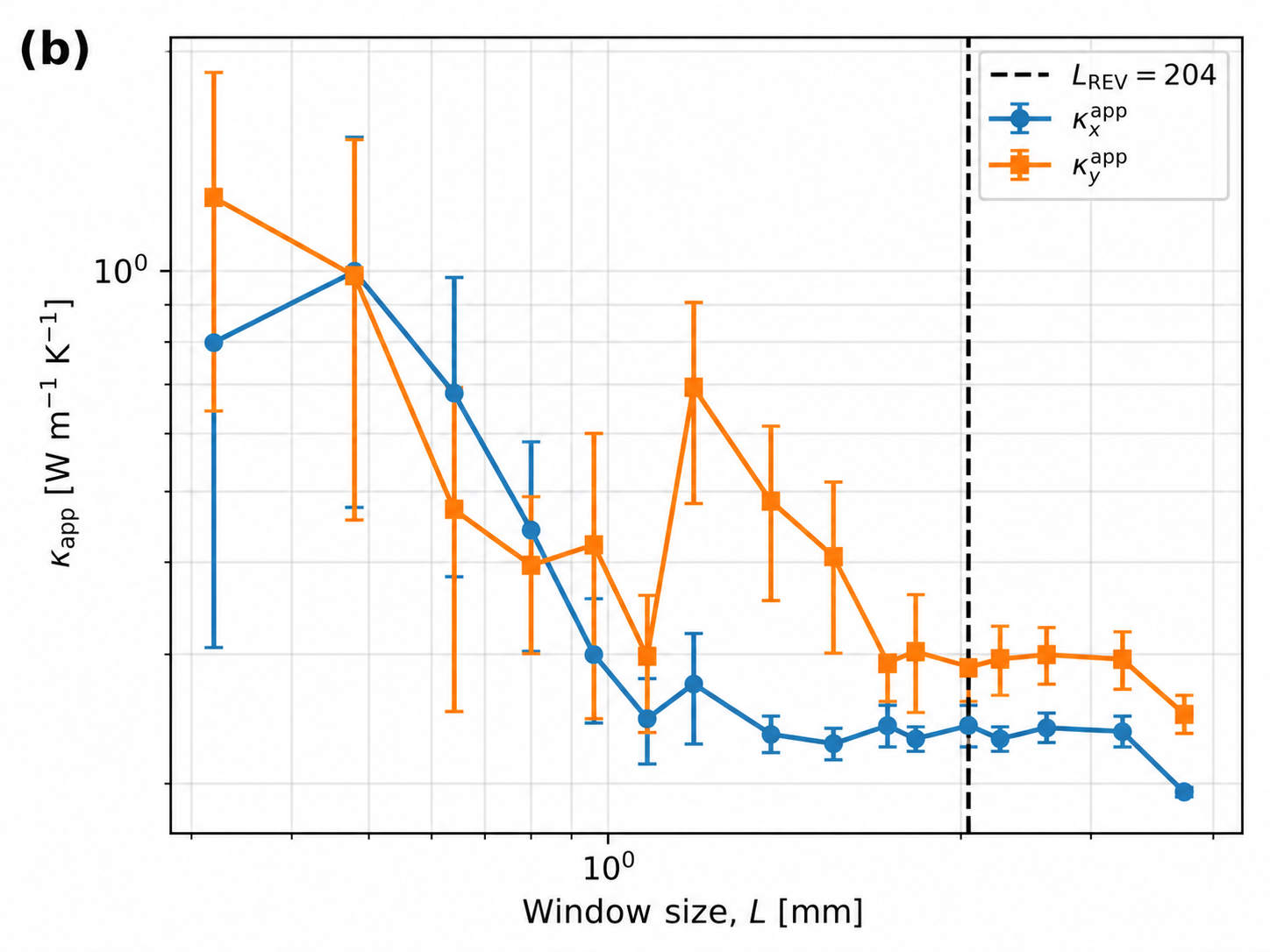}

    \vspace{0.5em}

    \includegraphics[width=0.45\linewidth]{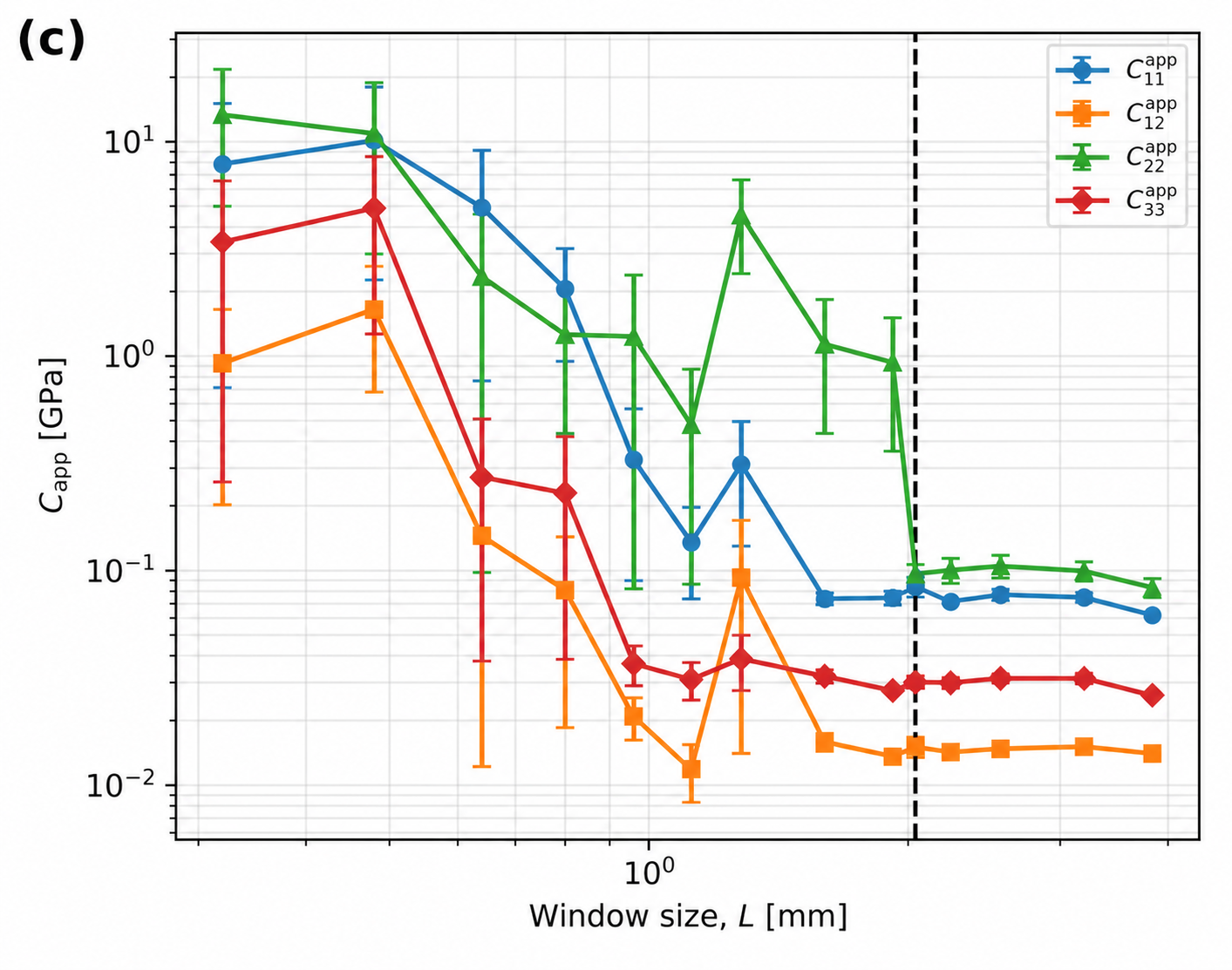}
    \includegraphics[width=0.48\linewidth]{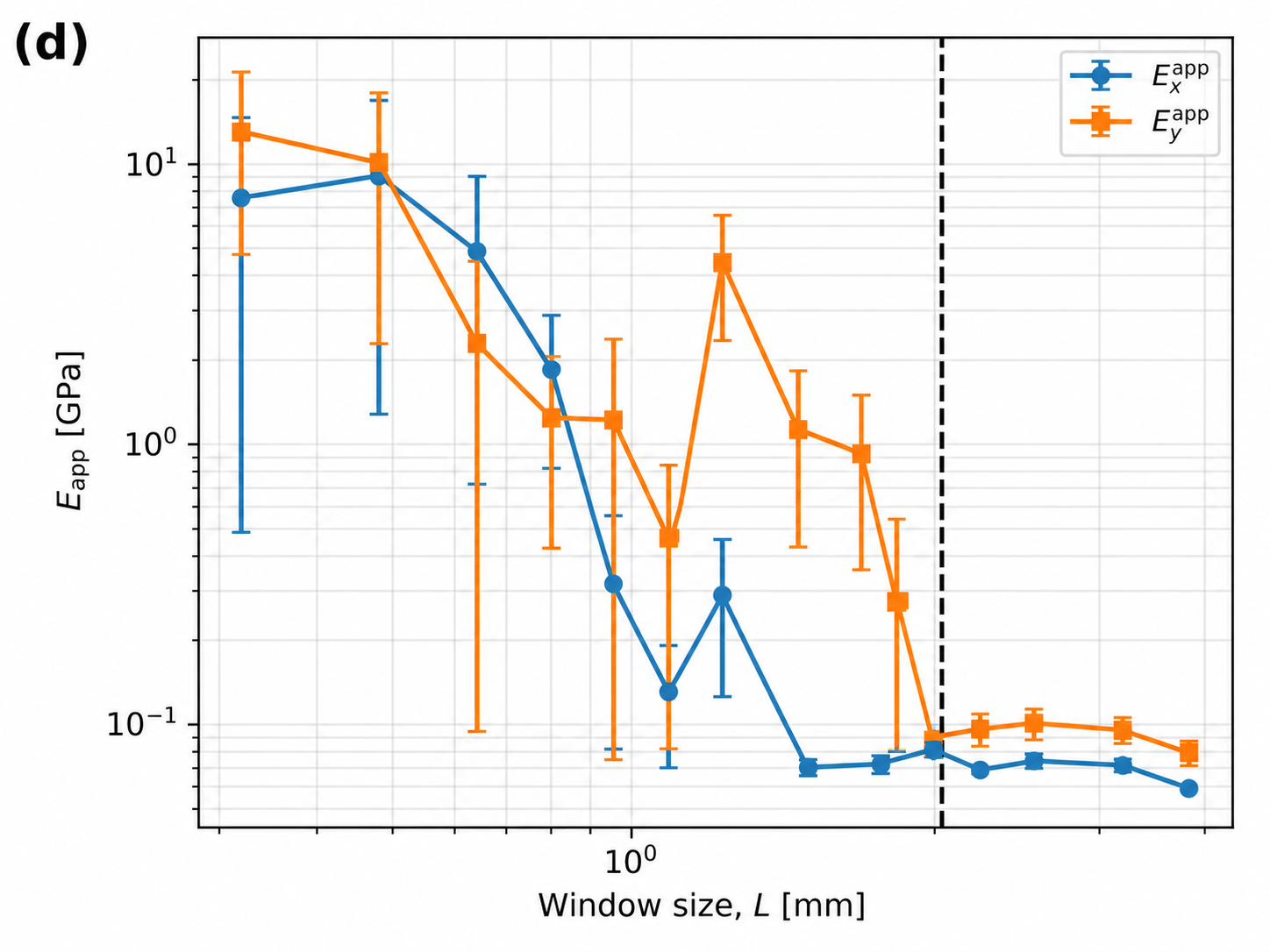}
    \caption{
    Independent physical validation of the BSE-derived structural REA.
    \textbf{(a)} Mineral/property composition used to construct the numerical
    fields \(\kappa(\mathbf{x})\), \(E(\mathbf{x})\), and \(\nu(\mathbf{x})\).
    The maps are dominated by quartz and pore/background pixels, with smaller
    contributions from feldspar, plagioclase, calcite, clay minerals, and
    accessory phases.
    \textbf{(b)} Apparent thermal-conductivity components
    \(\kappa_x^{\rm app}\) and \(\kappa_y^{\rm app}\), computed from the scalar
    periodic conduction problem.
    \textbf{(c)} Apparent elastic-stiffness components
    \(C_{11}^{\rm app}\), \(C_{12}^{\rm app}\), \(C_{22}^{\rm app}\), and
    \(C_{33}^{\rm app}\), computed from the local two-dimensional stiffness
    field constructed from \(E(\mathbf{x})\) and \(\nu(\mathbf{x})\).
    \textbf{(d)} Directional apparent Young moduli \(E_x^{\rm app}\) and
    \(E_y^{\rm app}\), obtained from the apparent compliance matrix. In panels
    \textbf{(b)}--\textbf{(d)}, the vertical dashed line marks the BSE-derived
    structural REA length converted to the QEMSCAN/property-map grid,
    \(L_{\rm REA}^{\rm prop}=204\) pixels. Error bars show the standard error of
    the accepted finite-window apparent responses pooled over stationary windows
    and image samples.
    }
    \label{fig:property_homogenization_summary}
\end{figure}

After accounting for the different spatial resolutions, the predicted support
corresponds to
\[
L_{\rm REA}^{\rm prop}=204\ \text{pixels},
\]
or approximately \(2.04~\mathrm{mm}\). This scale is indicated by the vertical
dashed line in Fig.~\ref{fig:property_homogenization_summary}. It is not inferred
from the property-convergence curves, but follows directly from the structural
REA through the grid conversion defined in
Eq.~\eqref{eq:Lrev_property_grid_general}.

At small support sizes, the apparent directional conductivities vary
substantially among individual windows
[Fig.~\ref{fig:property_homogenization_summary}(b)]. Each window samples a
different combination of mineral phases, pore fractions, and connected
heat-transfer pathways, leading to pronounced realization-to-realization
variability. As the window size approaches the structurally predicted support,
this variability decreases markedly and both directional conductivities enter
a weakly size-dependent regime.

The elastic response exhibits a consistent transition
[Fig.~\ref{fig:property_homogenization_summary}(c,d)]. At small support sizes,
the apparent stiffness components and directional Young's moduli depend
strongly on window position because the sampled domains contain incomplete
load-transfer pathways, isolated stiff-mineral clusters, and compliant
pore-rich regions. Near the transferred structural scale, the scatter is
substantially reduced and the elastic responses approach their corresponding
large-window regimes.

This agreement is nontrivial because the thermal and elastic fields are
constructed from mineral-resolved QEMSCAN maps and prescribed phase properties,
whereas the representative support is selected solely from the spatial
statistics of the BSE images. The stabilization of the constitutive responses
near the predicted scale therefore does not arise from fitting the structural
criterion to either property or from reusing the homogenized observables during
REA selection.

The results provide a two-dimensional physical validation of the structural
diagnostic. The support required to reproduce the long-wavelength organization
of the material also predicts the onset of weak size dependence in independent
thermal and elastic responses. Representativeness is therefore associated with
a structural scale having predictive relevance across multiple constitutive
observables, rather than with a length defined separately for each homogenized
property.

More generally, the same validation principle applies to volumetric material
fields. A structural REV determined from three-dimensional covariance
convergence should predict the stabilization of independently homogenized
three-dimensional constitutive tensors. The present study verifies this
principle in two dimensions, while its volumetric formulation is given in
Supplementary Note~S4.

\subsection{Why scalar plateaus can give false representativeness}

The results show that stabilization of a scalar observable is not a sufficient
criterion for representative support in granular materials. A finite domain can
reproduce a mean gray level, phase fraction, porosity, or even an apparent
property while still failing to reproduce the long-range organization of
particles, pores, and compositional domains. In the dune-sand images, the
high-wavenumber covariance content becomes similar at comparatively small
supports, whereas the low-wavenumber contribution continues to evolve. The
slowest-converging structural modes therefore control the support required to
capture the complete spatial organization of the sampled granular population.

A scalar plateau can also be generated for the wrong reason. When windows from
statistically incompatible regions are mixed, increasing the support suppresses
local differences and drives the average toward an apparently stable value.
That apparent stabilization reflects averaging across populations rather than
convergence within one population. The statistical-admissibility mask therefore
has a distinct role from the spectral criterion: it defines where
representative sizing is meaningful before any scale is selected.

\subsection{Long-wavelength structure provides a cross-property representative scale}

The most important validation is that the structural REA is determined without
using conductivity or elasticity, yet the transferred physical length marks the
onset of weak size dependence in both classes of response. The thermal and
elastic fields are constructed independently from mineral-resolved QEMSCAN maps
and prescribed phase properties. Their convergence near the BSE-derived support
therefore cannot result from fitting the representative length to either
constitutive observable.

This agreement gives the low-wavenumber criterion a direct physical
interpretation. Long-wavelength organization controls whether a finite domain
contains sufficiently complete heat-transfer and load-transfer networks to
reproduce the large-support response. The predicted support is consequently a
structural scale with relevance across distinct effective properties rather
than a different empirically fitted length for every observable. For
multiphysics homogenization, this distinction is important because it allows
the representative domain to be selected before each constitutive problem is
solved.

\subsection{Generality across granular materials and implications for geomechanics}

Although the validation material is Arabian dune sand, the mathematical
criterion does not depend on dune mineralogy, on BSE contrast, or on a
sand-specific constitutive model. Its defining operations are the localization
of statistically compatible domains and the persistent convergence of
long-wavelength spatial statistics. The same logic can be applied to other
granular materials represented by physically meaningful scalar or indicator
fields, including soils, weakly cemented sediments, crushed rock, granular
composites, and other heterogeneous particulate media imaged by BSE, SEM--EDS,
QEMSCAN, X-ray tomography, focused-ion-beam tomography, or related techniques.

For geomechanics and image-based homogenization, the practical implication is
that the computational domain should not be selected only from the plateau of
the property that will later be homogenized. A structural criterion can instead
be used to establish whether the sampled support is admissible and sufficiently
large before expensive thermal, hydraulic, or mechanical calculations are
performed. This separates microstructural representativeness from
property-specific response and makes the selection procedure transferable
across multiphysics applications.

\subsection{Limitations and extension from REA to REV}

The present numerical validation is two-dimensional. The conductivity,
stiffness, and directional Young's moduli reported here are in-plane apparent
properties of polished sections and are not interpreted as bulk
three-dimensional material properties. The value \(2.01~\mathrm{mm}\) is
therefore the structural REA of the analysed dune-sand sections under the stated
resolution, sampling protocol, and convergence tolerances; it is not presented
as a universal intrinsic length for all granular materials.

The generalization to three-dimensional voxelized materials is given in
Supplementary Note~S4. In three dimensions, planar
windows become voxelized subvolumes and the radial spectral average becomes a
spherical-shell average, with an optional directional decomposition for
anisotropic structures. Direct material-specific validation of the volumetric
criterion will require three-dimensional tomographic data and independently
homogenized three-dimensional constitutive tensors.

\section{Conclusions}
\label{sec:conclusions}

We have introduced a representative-support criterion for granular materials
that separates three requirements that are usually conflated: statistical
admissibility of the sampled domain, convergence of its long-wavelength
structural organization, and independent validation against effective physical
properties. This hierarchy prevents representative-area or
representative-volume selection from being reduced to the first apparent
plateau of a chosen scalar observable.

The concept was tested on seven large-area BSE images of Arabian dune sands.
After excluding statistically incompatible regions, persistent convergence of
the apparent mean and low-wavenumber covariance spectrum identifies a
structural REA of \(1536\) pixels, corresponding to approximately
\(2.01~\mathrm{mm}\). The scale is determined entirely from image structure and
is not fitted to a thermal or mechanical response.

When this independently selected length is transferred to mineral-resolved
QEMSCAN property fields, without property-specific recalibration, it coincides
with the onset of weak size dependence in apparent thermal conductivity,
elastic stiffness, and directional Young's moduli. The same structural scale
therefore predicts the convergence of distinct constitutive responses.

The dune-sand results support a broader conclusion: representative support in
granular materials should be assigned only after the sampled domain is shown
to be statistically compatible and its longest relevant spatial correlations
have converged. The numerical validation reported here is two-dimensional, but
the criterion has a direct mathematical extension to voxelized
three-dimensional materials. This provides a reproducible route from resolved
granular microstructure to representative continuum-scale support without
defining the REA or REV from an arbitrarily selected scalar plateau.

\section*{Data and code availability}
The full-resolution BSE images, QEMSCAN-derived numerical property maps,
stationary-domain masks, reference outputs, and supporting files required to
reproduce the analyses reported in this study are publicly available on Zenodo
at \url{https://doi.org/10.5281/zenodo.21717612}. The custom Python scripts used
for stationary-domain detection, structural REA determination, and thermal and
elastic homogenization are available at
\url{https://github.com/Christian48596/representativeness-emerges}. The
repository README provides the complete computational workflow, software
requirements, parameter settings, and commands used to reproduce the reported
results and figures.

\section*{CRediT authorship contribution statement}
Christian Tantardini: Conceptualization, Methodology, Formal analysis,
Writing -- original draft, Writing -- review \& editing.
Fernando Alonso-Marroqu\'in: Conceptualization, Methodology, Formal analysis,
Writing -- original draft, Writing -- review \& editing.
Abdullah Alqubalee: Investigation, Resources.
Eduardo Garzanti: Supervision, Writing -- review \& editing.
All authors discussed the results, revised the manuscript, and approved the final version.

\section*{Declaration of competing interest}
The authors declare that they have no known competing financial interests or
personal relationships that could have appeared to influence the work reported
in this paper.



\bibliographystyle{elsarticle-num}
\bibliography{main,REV_sandstones}


\clearpage

\input{SI_03}

\end{document}

%% file: SI_03.tex
\section*{Supplementary Information}

\setcounter{figure}{0}
\renewcommand{\thefigure}{S\arabic{figure}}

\setcounter{table}{0}
\renewcommand{\thetable}{S\arabic{table}}

\setcounter{equation}{0}
\renewcommand{\theequation}{S\arabic{equation}}

\setcounter{section}{0}
\renewcommand{\thesection}{S\arabic{section}}

\section{Statistical-admissibility diagnostics and comparison with conventional stationarity tests}
The present procedure should be distinguished from unit-root tests developed
for one-dimensional time series. The augmented Dickey--Fuller test, the
Phillips--Perron test, and the KPSS test operate on an ordered scalar sequence
and distinguish particular forms of unit-root, difference, or trend
stationarity
\cite{DickeyFuller1979,SaidDickey1984,PhillipsPerron1988,
KwiatkowskiEtAl1992}.
They are therefore not direct alternatives to the present spatial-domain
criterion. Applying such tests to a two-dimensional image would require an
externally imposed one-dimensional ordering, such as raster scanning, or
separate tests along selected rows and columns. The result would then depend
on the chosen traversal or direction and would not define a connected
two-dimensional material domain.

The Wiener--Khinchin consistency test provides a more closely related
second-order stationarity diagnostic
\cite{AriasCalluariEtAl2022,TangEtAl2024Stationarity}. In this test, the
autocovariance is first estimated directly from the mean-centred field in real
space and then Fourier transformed. Independently, the power spectral density
is estimated from the squared amplitude of the Fourier transform of the same
field. For a wide-sense stationary process, these two independently evaluated
spectral representations must agree, subject to consistent normalization and
finite-window conventions. Their agreement therefore provides a necessary,
but not sufficient, condition for second-order stationarity.

The Wiener--Khinchin test nevertheless addresses a different requirement from
the present stationary-domain procedure. It evaluates second-order consistency
within a prescribed observation domain, but does not by itself identify which
connected regions of a heterogeneous image belong to a common material
population. The present method first constructs a spatial admissibility mask
from local first- and second-moment compatibility and subsequently determines
the representative support from persistent low-wavenumber covariance
convergence. The Wiener--Khinchin test may therefore provide a complementary
second-order consistency diagnostic, but it does not replace either the
spatial-domain mask or the support-size convergence criterion used here.

The present framework is instead designed for the geometrical and physical
requirements of image-based representativeness analysis. It first identifies
connected spatial domains whose local first and second marginal moments are
compatible with robust image-level references. It then tests, independently,
whether the low-wavenumber covariance content of those domains converges with
increasing observation support. The local mask and the spectral-convergence
criterion are therefore complementary: the former establishes statistical
admissibility for spatial sampling, whereas the latter determines the support
required to reproduce long-wavelength structural organization. The procedure
is not proposed as a universal replacement for time-series stationarity tests,
but as a spatially invariant and purpose-specific criterion for finite-window
REA and REV analysis.

\begin{table}[H]
\begin{singlespace}
\centering
\caption{Comparison of conventional stationarity diagnostics with the spatial-admissibility criterion used in this work. ADF denotes the augmented Dickey--Fuller test, PP the Phillips--Perron test, and WK the Wiener--Khinchin consistency diagnostic. The key distinction is whether the procedure can localize an admissible material domain before support-size convergence is assessed.}
\label{tab:stationarity_method_comparison}
\footnotesize
\setlength{\tabcolsep}{4pt}
\renewcommand{\arraystretch}{1.17}
\begin{tabularx}{\textwidth}{@{}>{\bfseries\raggedright\arraybackslash}p{0.075\textwidth}
>{\raggedright\arraybackslash}p{0.18\textwidth}
>{\raggedright\arraybackslash}p{0.22\textwidth}
>{\raggedright\arraybackslash}X@{}}
\toprule
Approach & Input object & Principal diagnostic & Relation to spatial admissibility \\
\midrule
ADF &
Ordered 1-D sequence &
Autoregressive unit root &
No spatial localization; applying it to an image requires an imposed one-dimensional ordering. \\
\addlinespace[2pt]
PP &
Ordered 1-D sequence &
Unit root with nonparametric correction &
No spatial localization; it tests the imposed sequence rather than connected image regions. \\
\addlinespace[2pt]
KPSS &
Ordered 1-D sequence &
Level or trend stationarity &
Tests a time-series stationarity null and does not construct an admissible material domain. \\
\addlinespace[2pt]
WK &
Time series or random field &
Covariance--spectral consistency &
Useful second-order consistency check within a prescribed domain, but it does not identify which connected regions belong to one population. \\
\midrule
This work &
Spatial material field (2-D/3-D) &
Local-moment compatibility plus persistent low-$k$ convergence &
\textbf{Direct spatial output:} constructs the admissible mask and then determines the structural REA/REV within that domain. \\
\bottomrule
\end{tabularx}
\end{singlespace}
\end{table}

\section{QEMSCAN segmentation, constitutive mapping, and full numerical parameters}

\subsection{RGB-based mineral classification and property assignment}

The QEMSCAN outputs were represented as color-coded mineral maps on a regular
pixel grid. Each pixel was assigned to the closest reference color in the
mineral legend using the prescribed RGB-distance tolerance adopted in the
processing workflow. Pixels outside that tolerance were assigned to the unknown
class. The resulting mineral-index map was then converted into the thermal and
elastic fields used in the main manuscript.

\begin{figure}[t]
  \centering
  \includegraphics[width=\linewidth]{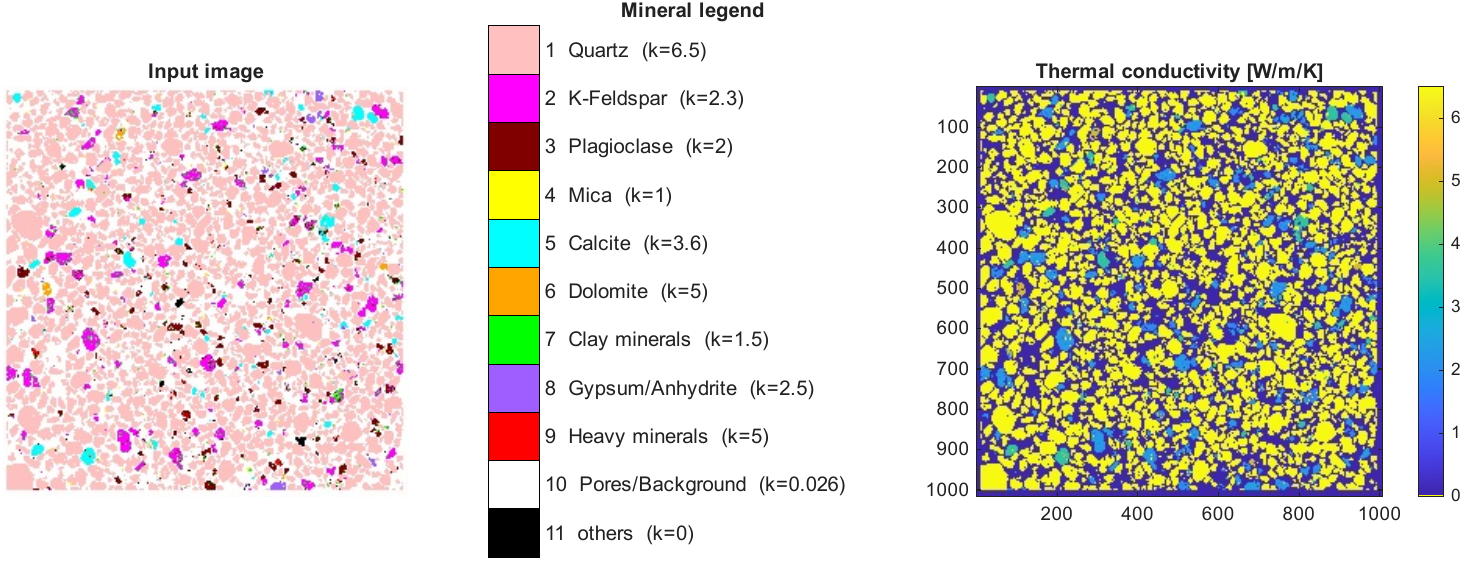}
  \caption{
  Conversion of a chemically mapped granular image into voxelized property
  fields.
  \textbf{(a)} Input QEMSCAN-derived mineral map represented as an RGB image.
  \textbf{(b)} Mineral legend used as the reference color code for phase
  assignment.
  \textbf{(c)} Example voxelized thermal-conductivity field
  \(\kappa(\mathbf{x})\). Each pixel is assigned to the closest legend color
  within a prescribed tolerance and then mapped to a mineral property. The same
  mineral-index map is used to construct \(E(\mathbf{x})\) and
  \(\nu(\mathbf{x})\). Pore pixels are assigned air properties for thermal
  conduction and a small nonzero stiffness for numerical stability. Unmatched
  pixels are treated as unknown and assigned fallback values.
  }
  \label{fig:S_qemscan_conversion}
\end{figure}

\begin{table}[H]
\begin{singlespace}
\centering
\caption{Reference mineral classes and constitutive values used to construct
the QEMSCAN-derived numerical property fields. The RGB triplets are the
classification colors; the constitutive values are fixed assignments used for
all windows and image samples.}
\label{tab:S_mineral_database}
\small
\setlength{\tabcolsep}{4.5pt}
\renewcommand{\arraystretch}{1.14}
\begin{tabular}{@{}r l
S[table-format=3.0] S[table-format=3.0] S[table-format=3.0]
S[table-format=1.3] S[table-format=3.3] S[table-format=1.2]@{}}
\toprule
& & \multicolumn{3}{c}{RGB reference} & \multicolumn{3}{c}{Assigned local properties} \\
\cmidrule(lr){3-5}\cmidrule(l){6-8}
ID & Mineral & {$R$} & {$G$} & {$B$} & {$\kappa$} & {$E$} & {$\nu$} \\
& & & & & {\scriptsize W\,m$^{-1}$\,K$^{-1}$} & {\scriptsize GPa} & {} \\
\midrule
1  & Quartz           & 255 & 192 & 192 & 6.5   & 94    & 0.08 \\
2  & K-Feldspar       & 255 & 0   & 255 & 2.3   & 70    & 0.25 \\
3  & Plagioclase      & 128 & 0   & 0   & 2.0   & 76    & 0.26 \\
4  & Mica             & 255 & 255 & 0   & 1.0   & 60    & 0.25 \\
5  & Calcite          & 0   & 255 & 255 & 3.6   & 72    & 0.31 \\
6  & Dolomite         & 255 & 165 & 0   & 5.0   & 94    & 0.28 \\
7  & Clay minerals    & 0   & 255 & 0   & 1.5   & 20    & 0.30 \\
8  & Gypsum/Anhydrite & 159 & 94  & 255 & 2.5   & 45    & 0.30 \\
9  & Heavy minerals   & 255 & 0   & 0   & 5.0   & 100   & 0.25 \\
10 & Pores/Background & 255 & 255 & 255 & 0.026 & 0.001 & 0.20 \\
11 & Others/unknown   & 0   & 0   & 0   & 0.0   & 0.001 & 0.20 \\
\bottomrule
\end{tabular}
\end{singlespace}
\end{table}

\subsection{Complete BSE support sequence and numerical parameter set}

For the full-resolution BSE analysis, the tested support sequence was
\begin{equation}
\mathcal L=\{L_n\}_{n=1}^{15},
\qquad
L_n=
\begin{cases}
128(n+1), & 1\le n\le 7,\\
256(n-3), & 8\le n\le 11,\\
512(n-7), & 12\le n\le 15,
\end{cases}
\quad \mathrm{pixels}.
\label{eq:S_BSE_tested_support_sequence}
\end{equation}

The remaining implementation parameters are summarized below.
\begin{singlespace}
\small
\setlength{\tabcolsep}{5pt}
\renewcommand{\arraystretch}{1.10}
\begin{longtable}{@{}>{\raggedright\arraybackslash}p{0.22\textwidth}
>{\raggedright\arraybackslash}p{0.24\textwidth}
>{\raggedright\arraybackslash}p{0.48\textwidth}@{}}
\caption{Numerical settings used in the final statistical-admissibility,
structural-REA, and property-validation analyses. Parameters are grouped by
the stage of the workflow in which they enter.}
\label{tab:S_numerical_parameters}\\
\toprule
Parameter & Value & Function in the analysis \\
\midrule
\endfirsthead
\multicolumn{3}{l}{\small\itshape Table~S3 continued}\\[2pt]
\toprule
Parameter & Value & Function in the analysis \\
\midrule
\endhead
\midrule
\multicolumn{3}{r}{\small\itshape Continued on next page}\\
\endfoot
\bottomrule
\endlastfoot
\multicolumn{3}{@{}l}{\textbf{Statistical-admissibility screening}} \\
\addlinespace[2pt]
$\tau_\mu$ & $0.02$ & Relative tolerance for the local stationarity-window mean. \\
$\tau_\sigma$ & $0.06$ & Relative tolerance for the local stationarity-window standard deviation. \\
$f_{\rm stat}$ & $0.98$ & Minimum fraction of a candidate REA window required to lie inside the admissible mask. \\
$R_{\rm stat}$ & $2048$ px & Side length of the windows used to construct the admissibility mask. \\
Stationarity stride & $512$ px & Translation step between successive stationarity windows. \\
\addlinespace[5pt]
\multicolumn{3}{@{}l}{\textbf{Persistent mean--spectral REA criterion}} \\
\addlinespace[2pt]
$\tau_Z$ & $0.025$ & Tolerance for the persistent apparent-mean residual. \\
$\tau_{\widehat C}$ & $0.30$ & Tolerance for the low-wavenumber covariance-spectrum residual. \\
$\epsilon_{\rm rel}$ & $10^{-15}$ & Relative numerical floor in the normalized spectral residual. \\
$n_{\rm med}$ & $5$ & Width of the running median applied to $r_Z(L)$. \\
$\mathcal K_{\rm low}$ & $k\le 0.25\,k_{\max}$ & Low-wavenumber band retained for structural convergence. \\
$w_j$ & $w_j^{\rm trap}/(q_j+\epsilon_k)$, $\epsilon_k=q_2$ & Inverse-wavenumber-weighted trapezoidal quadrature, emphasizing the longest resolved spatial fluctuations. \\
$J$ & $300$ & Number of points in the common radial wavenumber grid. \\
\addlinespace[5pt]
\multicolumn{3}{@{}l}{\textbf{Sampling, reference support, and physical validation}} \\
\addlinespace[2pt]
$N_{\rm window}$ (BSE) & Up to $50$ & Regular $7\times7$ candidate grid plus the centered window when distinct. \\
$N_{\rm window}$ (FEM) & $4$ & Maximum accepted windows evaluated per size in the property-level finite-element scan. \\
$\mathcal L$ & $256$--$4096$ px & BSE support range; the complete discrete sequence is given in Eq.~\eqref{eq:S_BSE_tested_support_sequence}. \\
$L_{\rm ref}$ & $4096$ px & Largest finite observational reference support. \\
Tail-valid fraction & $0.5$ & Minimum fraction of the non-reference larger-support tail required for a valid persistence decision. \\
\end{longtable}
\end{singlespace}

\section{Property-field preprocessing, finite-element implementation, and robustness checks}

\subsection{Ensemble reproducibility of the structural REA}
Ensemble reproducibility is evaluated after the REA has been selected. It is
used as an auxiliary diagnostic, not as a hard acceptance constraint. The
purpose is to verify that the selected support is not controlled by one
particular BSE image or one particular window location. The reproducibility
residual is
\begin{equation}
r_{\rm ens}(L)
=
\frac{
\operatorname{std}_{i=1,\ldots,N_s}
\left[
\langle Z\rangle_L^{(i)}
\right]
}{
\left|
\operatorname{mean}_{i=1,\ldots,N_s}
\left[
\langle Z\rangle_L^{(i)}
\right]
\right|+\epsilon
}.
\label{eq:raw_ensemble_residual}
\end{equation}
Here \(\langle Z\rangle_L^{(i)}\) is the image-level apparent gray-level mean at
support \(L\), \(N_s\) is the number of independent BSE samples, and
\(\epsilon\) prevents division by zero. Thus, \(r_{\rm ens}(L)\) measures the
relative spread of apparent means across the image ensemble.

The corresponding tail envelope is
\begin{equation}
\eta_{\rm ens}^{\rm tail}(L_m)
=
\max_{L_j\in\mathcal{L},\,L_j\ge L_m,\,L_j<L_{\rm ref}}
r_{\rm ens}(L_j).
\label{eq:eta_ens_tail}
\end{equation}
This quantity checks whether the apparent means remain reproducible across
independent samples over the larger-window tail. It is evaluated only after the
REA has been selected by the persistent mean--spectral rule in
the persistent mean--spectral rule defined in the main manuscript.

\subsection{Robust preprocessing of QEMSCAN-derived property fields}
For each image sample, a combined property field was constructed from the
conductivity, Young-modulus, and Poisson-ratio fields. The calculations used the
numerical QEMSCAN-derived property arrays directly, not rescaled visualization
images. Each property field was robustly standardized on its original numerical
scale. For a generic property
\(a_i(\mathbf{x})\), the median absolute deviation is defined as
\begin{equation}
\operatorname{MAD}[a_i]
=
\operatorname{median}_{\mathbf{x}}
\left[
\left|
a_i(\mathbf{x})
-
\operatorname{median}_{\mathbf{x}}a_i(\mathbf{x})
\right|
\right].
\label{eq:property_mad_definition}
\end{equation}
The robust scale is
\begin{equation}
s_{a,i}
=
\begin{cases}
c_{\rm MAD}\operatorname{MAD}[a_i],
&
\operatorname{MAD}[a_i]>0,
\\[1mm]
\operatorname{std}_{\mathbf{x}}[a_i],
&
\operatorname{MAD}[a_i]=0
\ \text{and}\
\operatorname{std}_{\mathbf{x}}[a_i]>0,
\\[1mm]
1,
&
\text{otherwise},
\end{cases}
\label{eq:robust_property_scale}
\end{equation}
where
\begin{equation}
c_{\rm MAD}
=
\frac{1}{\Phi^{-1}(3/4)}
\simeq
1.4826,
\label{eq:mad_consistency_factor}
\end{equation}
and \(\Phi^{-1}\) denotes the quantile function of the standard normal
distribution. The factor \(c_{\rm MAD}\) makes the scaled median absolute
deviation a Gaussian-consistent estimate of the standard deviation
\cite{RousseeuwCroux1993}. The standardized property field is consequently
\begin{equation}
z_{a,i}(\mathbf{x})
=
\frac{
a_i(\mathbf{x})
-
\operatorname{median}_{\mathbf{x}}a_i(\mathbf{x})
}{
s_{a,i}
}.
\label{eq:robust_property_standardization}
\end{equation}

The combined property-stationarity field is
\begin{equation}
Z_{{\rm prop},i}(\mathbf{x})
=
\left[
\frac{
z_{\kappa,i}^2(\mathbf{x})
+
z_{E,i}^2(\mathbf{x})
+
z_{\nu,i}^2(\mathbf{x})
}{3}
\right]^{1/2}.
\label{eq:combined_property_stationary_field}
\end{equation}
The same local mean--standard-deviation stationarity test used for the BSE
field is then applied to \(Z_{{\rm prop},i}\). This gives stationary masks on
the same grid as the property fields.

\subsection{Detailed periodic finite-element formulation}

The main manuscript gives the governing periodic cell problems and the
macroscopic definitions of the apparent tensors. The implementation used the
following explicit plane-strain and finite-element forms.

For thermal conduction, the scalar potential is decomposed as
\begin{equation}
u(\mathbf{x})
=
\mathbf E\cdot\mathbf{x}
+
\widetilde u(\mathbf{x}),
\qquad
\langle \widetilde u\rangle_{\Omega(L)}=0,
\label{eq:S_scalar_affine_periodic_methods}
\end{equation}
where \(\mathbf E\) is the imposed macroscopic gradient and
\(\widetilde u\) is periodic. The corrector problem is
\begin{equation}
-\nabla\cdot
\left[
\kappa(\mathbf{x})
\left(
\mathbf E+\nabla\widetilde u(\mathbf{x})
\right)
\right]
=0
\qquad
\text{in } \Omega(L).
\label{eq:S_scalar_cell_problem_methods}
\end{equation}
For the Cartesian loading \(\mathbf E^{(j)}=\widehat{\mathbf e}_j\), the
corresponding apparent-conductivity column is
\begin{equation}
\boldsymbol{\kappa}_{\rm app}^{(:,j)}(L)
=
\left\langle
\kappa(\mathbf{x})
\left[
\widehat{\mathbf e}_j+\nabla\widetilde u^{(j)}(\mathbf{x})
\right]
\right\rangle_{\Omega(L)}.
\label{eq:S_kappa_app_methods}
\end{equation}
The reported components are
\(\kappa_x^{\rm app}(L)=\kappa_{11}^{\rm app}(L)\) and
\(\kappa_y^{\rm app}(L)=\kappa_{22}^{\rm app}(L)\).

Under the plane-strain assumption, the local elastic properties were represented
in Voigt notation by
\begin{equation}
\mathbf C(\mathbf{x})
=
\frac{E(\mathbf{x})}
{\left[1+\nu(\mathbf{x})\right]
 \left[1-2\nu(\mathbf{x})\right]}
\begin{bmatrix}
1-\nu(\mathbf{x}) & \nu(\mathbf{x}) & 0\\
\nu(\mathbf{x}) & 1-\nu(\mathbf{x}) & 0\\
0 & 0 & \dfrac{1-2\nu(\mathbf{x})}{2}
\end{bmatrix}.
\label{eq:local_plane_strain_matrix}
\end{equation}

For elasticity, the displacement is decomposed as
\begin{equation}
\mathbf u(\mathbf{x})
=
\overline{\boldsymbol{\varepsilon}}\,\mathbf{x}
+
\widetilde{\mathbf u}(\mathbf{x}),
\qquad
\left\langle \widetilde{\mathbf u}\right\rangle_{\Omega(L)}
=
\mathbf 0,
\label{eq:S_elastic_affine_periodic_methods}
\end{equation}
where \(\overline{\boldsymbol{\varepsilon}}\) is the imposed macroscopic strain
and \(\widetilde{\mathbf u}\) is periodic. The elastic corrector problem is
\begin{equation}
-\nabla\cdot
\left\{
\mathbb{C}(\mathbf{x})
:
\left[
\overline{\boldsymbol{\varepsilon}}
+
\boldsymbol{\varepsilon}(\widetilde{\mathbf u})
\right]
\right\}
=
\mathbf 0
\qquad
\text{in } \Omega(L).
\label{eq:S_elastic_cell_problem_methods}
\end{equation}

In two dimensions, Voigt notation with engineering shear strain is used. The
three imposed strain states are
\begin{equation}
\overline{\boldsymbol{\varepsilon}}_{\rm v}^{(1)}
=
\begin{bmatrix}
1\\0\\0
\end{bmatrix},
\qquad
\overline{\boldsymbol{\varepsilon}}_{\rm v}^{(2)}
=
\begin{bmatrix}
0\\1\\0
\end{bmatrix},
\qquad
\overline{\boldsymbol{\varepsilon}}_{\rm v}^{(3)}
=
\begin{bmatrix}
0\\0\\1
\end{bmatrix}.
\label{eq:elastic_unit_strains_methods}
\end{equation}
The apparent stiffness matrix is obtained from the volume-averaged stress,
\begin{equation}
\mathbf C_{\rm app}^{(:,j)}(L)
=
\left\langle
\mathbf C(\mathbf{x})
\left[
\overline{\boldsymbol{\varepsilon}}_{\rm v}^{(j)}
+
\mathbf B(\mathbf{x})\mathbf d^{(j)}
\right]
\right\rangle_{\Omega(L)}.
\label{eq:S_Capp_methods}
\end{equation}
Here \(\mathbf B(\mathbf{x})\) is the finite-element strain--displacement
matrix and \(\mathbf d^{(j)}\) is the nodal displacement vector for the
\(j\)-th imposed strain state.
The
reported directional apparent Young moduli are obtained from the apparent
compliance matrix,
\begin{equation}
\mathbf S_{\rm app}(L)=\mathbf C_{\rm app}^{-1}(L),
\qquad
E_x^{\rm app}(L)=\frac{1}{S_{11}^{\rm app}(L)},
\qquad
E_y^{\rm app}(L)=\frac{1}{S_{22}^{\rm app}(L)}.
\label{eq:S_Eapp_methods}
\end{equation}

The scalar and elastic weak forms are discretized with bilinear quadrilateral
elements. Periodic degrees of freedom are identified on opposite faces of each
cell. The zero-mean constraints in
Eqs.~\eqref{eq:S_scalar_affine_periodic_methods} and
\eqref{eq:S_elastic_affine_periodic_methods} remove the constant scalar mode and
rigid translations.

For completeness, the discrete systems used in the implementation are given
below. For thermal conduction, multiplication of
Eq.~\eqref{eq:S_scalar_cell_problem_methods} by a periodic test function \(v\)
and integration by parts gives
\begin{equation}
\int_{\Omega(L)}
\kappa(\mathbf{x})\,
\nabla\widetilde u(\mathbf{x})\cdot\nabla v(\mathbf{x})\,d\Omega
=
-
\int_{\Omega(L)}
\kappa(\mathbf{x})\,
\mathbf E\cdot\nabla v(\mathbf{x})\,d\Omega .
\label{eq:S_scalar_weak_form}
\end{equation}
With the bilinear approximation
\begin{equation}
\widetilde u_h(\mathbf{x})
=
\sum_{n=1}^{N}u_n N_n(\mathbf{x}),
\label{eq:S_scalar_FE_expansion}
\end{equation}
the linear system is
\begin{equation}
\mathbf K\,\mathbf u=\mathbf f,
\label{eq:S_scalar_FE_system}
\end{equation}
where
\begin{align}
K_{mn}
&=
\int_{\Omega(L)}
\kappa(\mathbf{x})\,
\nabla N_n(\mathbf{x})\cdot\nabla N_m(\mathbf{x})\,d\Omega,
\label{eq:S_scalar_K}
\\
f_m
&=
-
\int_{\Omega(L)}
\kappa(\mathbf{x})\,
\mathbf E\cdot\nabla N_m(\mathbf{x})\,d\Omega .
\label{eq:S_scalar_f}
\end{align}
Because constant functions belong to the nullspace of the periodic scalar
operator, the zero-mean constraint is imposed in coefficient form as
\begin{equation}
\mathbf m^{\mathsf T}\mathbf u=0,
\qquad
m_n=
\int_{\Omega(L)}N_n(\mathbf{x})\,d\Omega ,
\label{eq:S_scalar_mean_constraint}
\end{equation}
leading to the augmented system
\begin{equation}
\begin{bmatrix}
\mathbf K & \mathbf m\\
\mathbf m^{\mathsf T} & 0
\end{bmatrix}
\begin{bmatrix}
\mathbf u\\
\lambda
\end{bmatrix}
=
\begin{bmatrix}
\mathbf f\\
0
\end{bmatrix}.
\label{eq:S_scalar_augmented_system}
\end{equation}
The microscopic flux is reconstructed as
\begin{equation}
\mathbf q_h(\mathbf{x})
=
-\kappa(\mathbf{x})
\left[
\mathbf E+\mathbf B_\nabla(\mathbf{x})\mathbf u
\right],
\label{eq:S_scalar_flux_reconstruction}
\end{equation}
where \(\mathbf B_\nabla\) collects the gradients of the scalar shape
functions. Volume averaging of this field gives the apparent conductivity
reported in the main manuscript.

For elasticity, the weak form of
Eq.~\eqref{eq:S_elastic_cell_problem_methods} is
\begin{equation}
\int_{\Omega(L)}
\boldsymbol{\varepsilon}(\mathbf v)
:
\mathbb C(\mathbf{x})
:
\boldsymbol{\varepsilon}(\widetilde{\mathbf u})
\,d\Omega
=
-
\int_{\Omega(L)}
\boldsymbol{\varepsilon}(\mathbf v)
:
\mathbb C(\mathbf{x})
:
\overline{\boldsymbol{\varepsilon}}
\,d\Omega .
\label{eq:S_elastic_weak_form}
\end{equation}
The periodic displacement fluctuation is approximated as
\begin{equation}
\widetilde{\mathbf u}_h(\mathbf{x})
=
\sum_{n=1}^{N}N_n(\mathbf{x})\,\mathbf d_n,
\label{eq:S_elastic_FE_expansion}
\end{equation}
and, using engineering-shear Voigt notation,
\begin{equation}
\boldsymbol{\varepsilon}_{\rm v}
(\widetilde{\mathbf u}_h)
=
\mathbf B(\mathbf{x})\,\mathbf d .
\label{eq:S_elastic_B_relation}
\end{equation}
For a Q4 element, the nodal strain--displacement block is
\begin{equation}
\mathbf B_n(\mathbf{x})
=
\begin{bmatrix}
N_{n,x} & 0\\
0 & N_{n,y}\\
N_{n,y} & N_{n,x}
\end{bmatrix}.
\label{eq:S_Q4_B_block}
\end{equation}
The discrete elastic system is
\begin{equation}
\mathbf K\,\mathbf d=\mathbf f,
\label{eq:S_elastic_FE_system}
\end{equation}
with
\begin{align}
\mathbf K
&=
\int_{\Omega(L)}
\mathbf B^{\mathsf T}(\mathbf{x})
\mathbf C(\mathbf{x})
\mathbf B(\mathbf{x})\,d\Omega ,
\label{eq:S_elastic_K}
\\
\mathbf f
&=
-
\int_{\Omega(L)}
\mathbf B^{\mathsf T}(\mathbf{x})
\mathbf C(\mathbf{x})
\overline{\boldsymbol{\varepsilon}}_{\rm v}\,d\Omega .
\label{eq:S_elastic_f}
\end{align}
The two translational null modes are removed by imposing
\begin{equation}
\mathbf G\,\mathbf d=\mathbf 0,
\label{eq:S_elastic_mean_constraint}
\end{equation}
where \(\mathbf G\) contains the integrals of the shape functions for each
displacement component. The constrained problem is therefore
\begin{equation}
\begin{bmatrix}
\mathbf K & \mathbf G^{\mathsf T}\\
\mathbf G & \mathbf 0
\end{bmatrix}
\begin{bmatrix}
\mathbf d\\
\boldsymbol{\lambda}
\end{bmatrix}
=
\begin{bmatrix}
\mathbf f\\
\mathbf 0
\end{bmatrix}.
\label{eq:S_elastic_augmented_system}
\end{equation}
After solution, the microscopic strain and stress are reconstructed as
\begin{align}
\boldsymbol{\varepsilon}_{{\rm v},h}(\mathbf{x})
&=
\overline{\boldsymbol{\varepsilon}}_{\rm v}
+
\mathbf B(\mathbf{x})\mathbf d,
\label{eq:S_elastic_strain_reconstruction}
\\
\boldsymbol{\sigma}_{{\rm v},h}(\mathbf{x})
&=
\mathbf C(\mathbf{x})
\left[
\overline{\boldsymbol{\varepsilon}}_{\rm v}
+
\mathbf B(\mathbf{x})\mathbf d
\right].
\label{eq:S_elastic_stress_reconstruction}
\end{align}
Volume averaging of the stress for the three independent imposed strain states
gives the columns of \(\mathbf C_{\rm app}(L)\).

The resulting constrained linear systems are solved directly for each window
and loading. Thus, the apparent tensors are not affected by the fixed-point,
Krylov, or acceleration tolerances that enter many FFT-based homogenization
algorithms \cite{MoulinecSuquet1998,Zeman2010,Schneider2021}.

\subsection{Auxiliary consistency figure}
\begin{figure}[H]
    \centering
    \includegraphics[width=\linewidth]{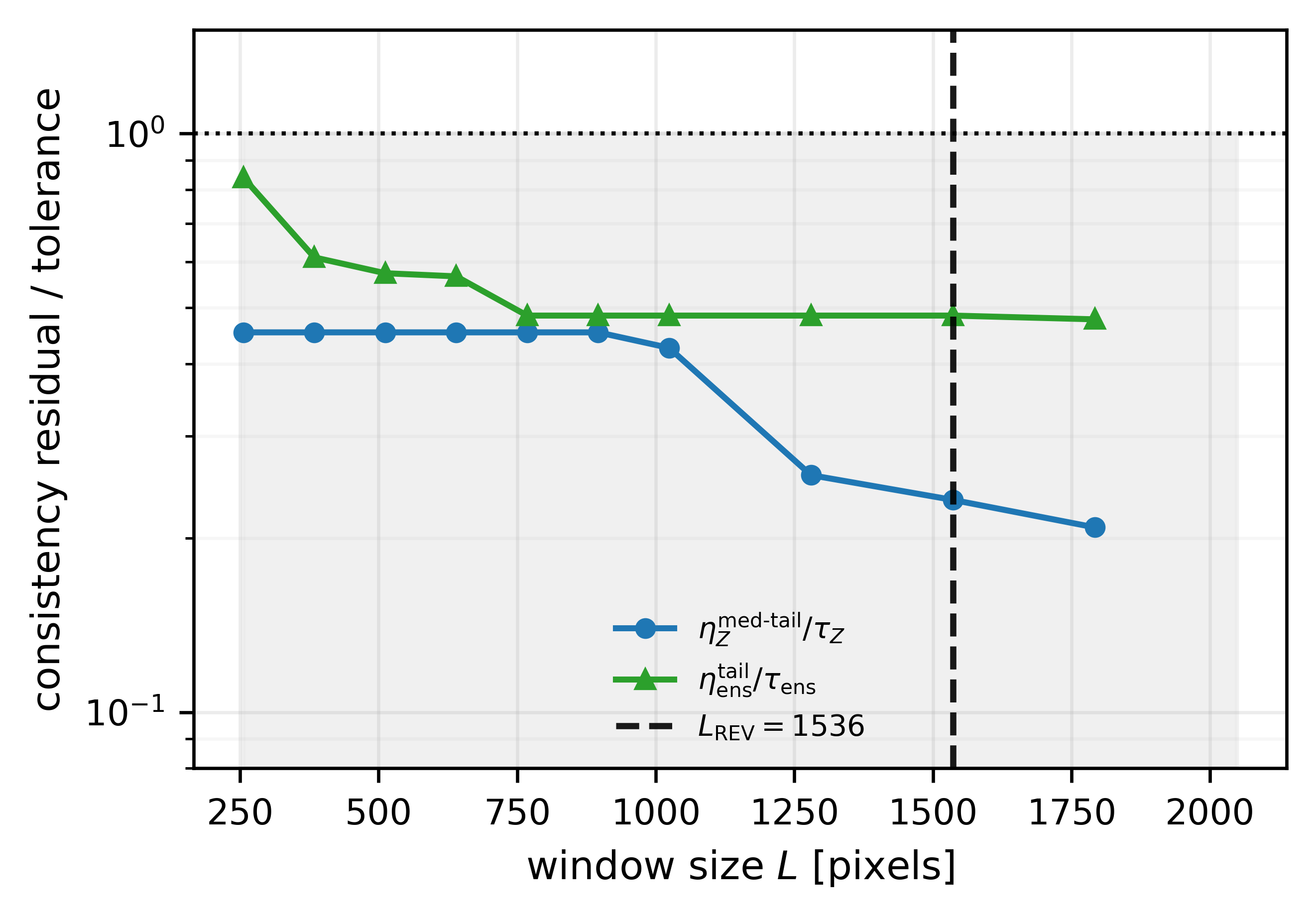}
    \caption{
    Consistency checks for the stationary-domain BSE REA analysis.
    The blue curve shows the normalized median-tail residual of the ensemble
    apparent mean, \(\eta_Z^{\rm med-tail}/\tau_Z\), which measures whether the
    gray-level mean remains stable over the larger-window tail. The green curve
    shows the ensemble-reproducibility residual,
    \(\eta_{\rm ens}^{\rm tail}\), which measures the relative spread of
    apparent gray-level means across the seven independent BSE samples. These
    quantities are reported as auxiliary diagnostics, not as additional
    acceptance constraints. The structural REA is selected by the persistent
    mean--spectral rule defined in the main manuscript; the consistency
    checks verify that the selected support is not controlled by one image or
    one window location.
    }
    \label{fig:S_rev_consistency_checks}
\end{figure}

\section{Three-dimensional extension to voxelized granular materials}
\label{subsec:three_dimensional_extension}
The numerical analysis presented in this work is performed on two-dimensional
BSE and QEMSCAN fields. Nevertheless, the statistical-admissibility and
structural-convergence framework is not intrinsically dimension dependent and
admits a direct mathematical generalization to three-dimensional voxelized
images. We provide this formulation to define the volumetric counterpart of
the present analysis and to clarify its applicability to tomographic datasets.
The equations below establish the three-dimensional form of the criterion; no
three-dimensional numerical validation or material-specific three-dimensional
REA value is claimed in the present study.

Let \(\Omega_i^{(3)}\subset\mathbb{R}^{3}\) denote the voxelized domain of
the \(i\)-th independent image volume, and let
\[
Z_i^{(3)}
\in
L^2\!\left(\Omega_i^{(3)}\right)
\]
denote the corresponding scalar image field. Depending on the imaging
modality, \(Z_i^{(3)}\) may represent gray level, X-ray attenuation, density,
porosity, or another continuous local material descriptor. For categorical
multiphase images, the analysis is applied to phase-indicator fields rather
than to arbitrary numerical phase labels.

Throughout this subsection, spatial coordinates are expressed in voxel-index
units. Accordingly, \(R_{\rm stat}\), \(L\), and \(L_{\rm ref}\) denote edge
lengths measured in voxels. The integral expressions provide the continuum
form of the framework; in a voxelized implementation, they are replaced by
the corresponding discrete sums. Physical edge lengths are recovered below
from the voxel spacings \(\Delta x_1\), \(\Delta x_2\), and \(\Delta x_3\).

For a phase \(\beta\), the corresponding indicator field is
\begin{equation}
I_{i,\beta}^{(3)}(\mathbf{x})
=
\begin{cases}
1,
&
\mathbf{x}\ \text{belongs to phase }\beta,
\\
0,
&
\text{otherwise}.
\end{cases}
\label{eq:three_dimensional_phase_indicator}
\end{equation}
The convergence criterion may then be imposed separately on each relevant
phase or jointly on a prescribed collection of indicator fields.

For a stationarity window of side length \(R_{\rm stat}\), centred at
\(\mathbf{x}=(x_1,x_2,x_3)\in\Omega_i^{(3)}\), define the cubic subdomain
\begin{equation}
W_{R_{\rm stat}}^{(i,3)}(\mathbf{x})
=
\left\{
\mathbf{y}\in\Omega_i^{(3)}
:
\left|y_\alpha-x_\alpha\right|
\le
\frac{R_{\rm stat}}{2},
\quad
\alpha=1,2,3
\right\}.
\label{eq:three_dimensional_stationarity_window}
\end{equation}
Thus, \(W_{R_{\rm stat}}^{(i,3)}(\mathbf{x})\) is the portion of the cube
centred at \(\mathbf{x}\) that lies inside the available image volume.

The local mean and local standard deviation are
\begin{align}
\mu_i^{(3)}(\mathbf{x})
&=
\frac{1}{
\left|
W_{R_{\rm stat}}^{(i,3)}(\mathbf{x})
\right|
}
\int_{W_{R_{\rm stat}}^{(i,3)}(\mathbf{x})}
Z_i^{(3)}(\mathbf{y})\,d\mathbf{y},
\label{eq:three_dimensional_local_mean}
\\
\sigma_i^{(3)}(\mathbf{x})
&=
\left[
\frac{1}{
\left|
W_{R_{\rm stat}}^{(i,3)}(\mathbf{x})
\right|
}
\int_{W_{R_{\rm stat}}^{(i,3)}(\mathbf{x})}
\left[
Z_i^{(3)}(\mathbf{y})
-
\mu_i^{(3)}(\mathbf{x})
\right]^2
d\mathbf{y}
\right]^{1/2}.
\label{eq:three_dimensional_local_std}
\end{align}

Robust volume-level references are defined by
\begin{equation}
\mu_{{\rm med},i}^{(3)}
=
\operatorname{median}_{\mathbf{x}}
\mu_i^{(3)}(\mathbf{x}),
\qquad
\sigma_{{\rm med},i}^{(3)}
=
\operatorname{median}_{\mathbf{x}}
\sigma_i^{(3)}(\mathbf{x}),
\label{eq:three_dimensional_reference_medians}
\end{equation}
where the medians are evaluated over all valid stationarity-window centres.

The three-dimensional statistically admissible mask is
\begin{equation}
S_i^{(3)}(\mathbf{x})
=
\mathbbm{1}
\left[
\frac{
\left|
\mu_i^{(3)}(\mathbf{x})
-
\mu_{{\rm med},i}^{(3)}
\right|
}{
\left|
\mu_{{\rm med},i}^{(3)}
\right|
+
\epsilon
}
\le
\tau_\mu^{(3)}
\right]
\mathbbm{1}
\left[
\frac{
\left|
\sigma_i^{(3)}(\mathbf{x})
-
\sigma_{{\rm med},i}^{(3)}
\right|
}{
\left|
\sigma_{{\rm med},i}^{(3)}
\right|
+
\epsilon
}
\le
\tau_\sigma^{(3)}
\right],
\label{eq:three_dimensional_stationary_mask}
\end{equation}
where \(\mathbbm{1}[\cdot]\) is the indicator function. The corresponding
statistically admissible subvolume is
\begin{equation}
\Omega_{{\rm stat},i}^{(3)}
=
\left\{
\mathbf{x}
\in
\Omega_i^{(3)}
:
S_i^{(3)}(\mathbf{x})=1
\right\}.
\label{eq:three_dimensional_stationary_domain}
\end{equation}
As in the two-dimensional formulation, this test establishes local
compatibility of the first two marginal moments but does not by itself prove
weak stationarity. Convergence of the second-order spatial structure is tested
independently through the covariance spectrum.

For a candidate support of side length \(L\), centred at
\(\mathbf{x}_p=(x_{p,1},x_{p,2},x_{p,3})\), define
\begin{equation}
Q_{L,p}^{(i,3)}
=
\left\{
\mathbf{y}\in\Omega_i^{(3)}
:
\left|y_\alpha-x_{p,\alpha}\right|
\le
\frac{L}{2},
\quad
\alpha=1,2,3
\right\}.
\label{eq:three_dimensional_candidate_cube}
\end{equation}
The candidate voxel subvolume is retained only when
\begin{equation}
\frac{
\left|
Q_{L,p}^{(i,3)}
\cap
\Omega_{{\rm stat},i}^{(3)}
\right|
}{
\left|
Q_{L,p}^{(i,3)}
\right|
}
\ge
f_{\rm stat}^{(3)}.
\label{eq:three_dimensional_mask_coverage}
\end{equation}

For every accepted voxel subvolume, the apparent field mean is
\begin{equation}
\langle Z\rangle_{L,p}^{(i,3)}
=
\frac{1}{
\left|
Q_{L,p}^{(i,3)}
\right|
}
\int_{Q_{L,p}^{(i,3)}}
Z_i^{(3)}(\mathbf{x})\,d\mathbf{x}.
\label{eq:three_dimensional_apparent_mean}
\end{equation}
The image-level and ensemble-level apparent means are
\begin{equation}
\langle Z\rangle_L^{(i,3)}
=
\frac{1}{N_L^{(i,3)}}
\sum_{p=1}^{N_L^{(i,3)}}
\langle Z\rangle_{L,p}^{(i,3)},
\qquad
\overline Z_L^{(3)}
=
\frac{1}{N_s}
\sum_{i=1}^{N_s}
\langle Z\rangle_L^{(i,3)}.
\label{eq:three_dimensional_ensemble_mean}
\end{equation}
The corresponding apparent-mean residual is
\begin{equation}
r_Z^{(3)}(L)
=
\frac{
\left|
\overline Z_L^{(3)}
-
\overline Z_{L_{\rm ref}}^{(3)}
\right|
}{
\left|
\overline Z_{L_{\rm ref}}^{(3)}
\right|
+
\epsilon
}.
\label{eq:three_dimensional_mean_residual}
\end{equation}

For the covariance-spectral diagnostic, one admissible cube is selected
deterministically for each image and support size. Let
\(\mathcal P_L^{(i,3)}\) denote the set of accepted cube indices and let
\(\mathbf{x}_{c,i}\) be the centre of the complete image volume. We define
\begin{equation}
p_\star(i,L)
=
\operatorname*{arg\,min}_{p\in\mathcal P_L^{(i,3)}}
\left\|
\mathbf{x}_p-\mathbf{x}_{c,i}
\right\|_2,
\label{eq:three_dimensional_selected_cube_index}
\end{equation}
with a fixed deterministic rule used to resolve possible ties. The selected
spectral cube is
\begin{equation}
Q_{L,\star}^{(i,3)}
=
Q_{L,p_\star(i,L)}^{(i,3)}.
\label{eq:three_dimensional_selected_cube}
\end{equation}
Its local mean is
\begin{equation}
\mu_{L,\star}^{(i,3)}
=
\frac{1}{
\left|
Q_{L,\star}^{(i,3)}
\right|
}
\int_{Q_{L,\star}^{(i,3)}}
Z_i^{(3)}(\mathbf{x})\,d\mathbf{x}.
\label{eq:three_dimensional_selected_cube_mean}
\end{equation}

The mean-centred field within the selected spectral cube is
\begin{equation}
\widetilde Z_{L,\star}^{(i,3)}(\mathbf{x})
=
Z_i^{(3)}(\mathbf{x})
-
\mu_{L,\star}^{(i,3)}.
\label{eq:three_dimensional_mean_centered_field}
\end{equation}
For a displacement vector \(\mathbf{r}\), define the overlap domain
\begin{equation}
Q_{L,\star}^{(i,3)}(\mathbf{r})
=
\left\{
\mathbf{x}\in Q_{L,\star}^{(i,3)}
:
\mathbf{x}+\mathbf{r}
\in
Q_{L,\star}^{(i,3)}
\right\}.
\label{eq:three_dimensional_overlap_domain}
\end{equation}
The corresponding set of admissible displacement vectors is
\begin{equation}
\mathcal R_{L,\star}^{(i,3)}
=
\left\{
\mathbf{r}\in\mathbb{R}^{3}
:
\left|
Q_{L,\star}^{(i,3)}(\mathbf{r})
\right|
>0
\right\}.
\label{eq:three_dimensional_admissible_displacements}
\end{equation}
The finite-volume autocovariance is evaluated directly from the image field as
\begin{equation}
C_L^{(i,3)}(\mathbf{r})
=
\frac{1}{
\left|
Q_{L,\star}^{(i,3)}(\mathbf{r})
\right|
}
\int_{Q_{L,\star}^{(i,3)}(\mathbf{r})}
\widetilde Z_{L,\star}^{(i,3)}(\mathbf{x})
\widetilde Z_{L,\star}^{(i,3)}(\mathbf{x}+\mathbf{r})
\,d\mathbf{x},
\qquad
\mathbf{r}\in\mathcal R_{L,\star}^{(i,3)}.
\label{eq:three_dimensional_autocovariance}
\end{equation}
The corresponding covariance spectrum is defined as the Fourier transform of
this directly evaluated autocovariance,
\begin{equation}
\widehat C_L^{(i,3)}(\mathbf{k})
=
\int_{\mathcal R_{L,\star}^{(i,3)}}
C_L^{(i,3)}(\mathbf{r})
\exp\left(
-\mathrm{i}\mathbf{k}\cdot\mathbf{r}
\right)
\,d\mathbf{r},
\qquad
\mathbf{k}=(k_1,k_2,k_3).
\label{eq:three_dimensional_covariance_spectrum}
\end{equation}
In a voxelized implementation, the autocovariance is evaluated directly on
the discrete displacement grid, and its Fourier transform is computed using
a three-dimensional discrete Fourier transform.

For an approximately isotropic material, the discrete Fourier modes are
grouped into spherical shells
\begin{equation}
\mathcal S_j
=
\left\{
\mathbf{k}_\nu\ne\mathbf{0}
:
q_{j-\frac12}
\le
\left|
\mathbf{k}_\nu
\right|
<
q_{j+\frac12}
\right\},
\label{eq:three_dimensional_fourier_shells}
\end{equation}
and the shell-averaged spectrum is
\begin{equation}
\widehat C_L^{(i,3)}(q_j)
=
\frac{1}{
\left|
\mathcal S_j
\right|
}
\sum_{\mathbf{k}_\nu\in\mathcal S_j}
\widehat C_L^{(i,3)}(\mathbf{k}_\nu).
\label{eq:three_dimensional_shell_average}
\end{equation}
The zero-wavenumber value is excluded from the structural residual, which is
defined only over nonzero spatial modes. Although the image field is
mean-centred before the autocovariance is evaluated, the zero-wavenumber value
of its finite-window Fourier transform is not generally identically zero when
a displacement-dependent overlap normalization is used. The radial grid
therefore contains only positive wavenumbers,
\begin{equation}
0<q_1<q_2<\cdots<q_J\le k_{\max}^{(3)}.
\label{eq:three_dimensional_positive_radial_grid}
\end{equation}
For voxel spacings \(\Delta x_\alpha\), the largest direction-independent
radial band that remains below the Nyquist limit in every coordinate direction
is bounded by
\begin{equation}
k_{\max}^{(3)}
=
\min_{\alpha=1,2,3}
\frac{\pi}{\Delta x_\alpha}.
\label{eq:three_dimensional_nyquist_bound}
\end{equation}
For isotropic voxels, this reduces to
\(k_{\max}^{(3)}=\pi/\Delta x\).

The low-wavenumber weighted norm is
\begin{equation}
\left\|
f
\right\|_{w,\mathcal K_{\rm low}}^{(3)}
=
\left[
\sum_{q_j\in\mathcal K_{\rm low}}
w_j f^2(q_j)
\right]^{1/2},
\label{eq:three_dimensional_weighted_lowk_norm}
\end{equation}
with
\begin{equation}
w_j
=
\frac{
w_j^{\rm trap}
}{
q_j+\epsilon_k
},
\qquad
\epsilon_k=q_1.
\label{eq:three_dimensional_lowk_weights}
\end{equation}
Because \(q_1\) is the smallest positive radial wavenumber, the denominator in
Eq.~\eqref{eq:three_dimensional_lowk_weights} is finite. The inverse-wavenumber
factor emphasizes the longest resolved spatial fluctuations after the
spherical-shell average has been performed.

A reference-spectrum-based numerical floor is defined by
\begin{equation}
\epsilon_{\widehat C}^{(3)}
=
\epsilon_{\rm rel}
\max_{s=1,\ldots,N_s}
\left\|
\widehat C_{L_{\rm ref}}^{(s,3)}
\right\|_{w,\mathcal K_{\rm low}}^{(3)},
\qquad
\epsilon_{\rm rel}=10^{-15}.
\label{eq:three_dimensional_spectral_floor}
\end{equation}
The value of \(\epsilon_{\rm rel}\) is a floating-point safeguard rather than
a physical or statistical tolerance. It affects the residual only in the
degenerate case in which the low-wavenumber norm of the reference spectrum is
numerically indistinguishable from zero.

The image-level low-wavenumber spectral residual is
\begin{equation}
\eta_{\widehat C}^{(i,3)}(L)
=
\frac{
\left\|
\widehat C_L^{(i,3)}
-
\widehat C_{L_{\rm ref}}^{(i,3)}
\right\|_{w,\mathcal K_{\rm low}}^{(3)}
}{
\max
\left[
\left\|
\widehat C_{L_{\rm ref}}^{(i,3)}
\right\|_{w,\mathcal K_{\rm low}}^{(3)},
\epsilon_{\widehat C}^{(3)}
\right]
},
\qquad
L<L_{\rm ref}.
\label{eq:three_dimensional_spectral_residual}
\end{equation}
The ensemble residual is
\begin{equation}
\eta_{\widehat C}^{(3)}(L)
=
\frac{1}{N_s}
\sum_{i=1}^{N_s}
\eta_{\widehat C}^{(i,3)}(L).
\label{eq:three_dimensional_ensemble_spectral_residual}
\end{equation}

After applying the same running-median regularization used for the
two-dimensional apparent-mean sequence, the persistent mean tail is
\begin{equation}
\eta_Z^{{\rm med-tail},(3)}(L_m)
=
\max_{
\substack{
L_j\in\mathcal L,\,
L_j\ge L_m,\\
L_j<L_{\rm ref}
}
}
\widetilde r_Z^{(3)}(L_j).
\label{eq:three_dimensional_mean_tail}
\end{equation}
The three-dimensional structural REV is defined as the smallest tested voxel
support for which both the apparent field mean and the low-wavenumber
covariance spectrum remain converged over the complete larger-support tail,
\begin{equation}
\begin{aligned}
L_{\rm REV}^{(3)}
=
\min\bigg\{
L_m\in\mathcal L,\;
L_m<L_{\rm ref}:
\quad&
\eta_Z^{{\rm med-tail},(3)}(L_m)
\le
\tau_Z^{(3)},
\\
&
\eta_{\widehat C}^{(3)}(L')
\le
\tau_{\widehat C}^{(3)}
\quad
\forall\,L'\in\mathcal L,
\\
&
L'\ge L_m,
\qquad
L'<L_{\rm ref}
\bigg\}.
\end{aligned}
\label{eq:three_dimensional_rev_rule}
\end{equation}

The three-dimensional tolerances
\(\tau_\mu^{(3)}\), \(\tau_\sigma^{(3)}\),
\(\tau_Z^{(3)}\), and \(\tau_{\widehat C}^{(3)}\)
retain the same operational meanings as their two-dimensional counterparts.
Their numerical specification is dataset dependent and belongs to the
application of the framework to a particular volumetric dataset; no
three-dimensional tolerance values are inferred from the present planar data.

For voxel spacings
\(\Delta x_1\), \(\Delta x_2\), and \(\Delta x_3\), the physical edge lengths
of a representative support containing \(L_{\rm REV}^{(3)}\) voxels along
each coordinate direction are
\begin{equation}
\ell_{{\rm REV},\alpha}^{(3)}
=
L_{\rm REV}^{(3)}
\Delta x_\alpha,
\qquad
\alpha=1,2,3.
\label{eq:three_dimensional_physical_lengths}
\end{equation}
The corresponding physical representative volume is
\begin{equation}
V_{\rm REV}
=
\prod_{\alpha=1}^{3}
\ell_{{\rm REV},\alpha}^{(3)}
=
\left[
L_{\rm REV}^{(3)}
\right]^3
\Delta x_1\Delta x_2\Delta x_3.
\label{eq:three_dimensional_physical_volume}
\end{equation}
When the voxel spacings are anisotropic, these edge lengths define a
rectangular cuboid in physical space. The physical support is cubic only when
\(\Delta x_1=\Delta x_2=\Delta x_3\).
For isotropic voxels,
\(\Delta x_1=\Delta x_2=\Delta x_3=\Delta x\), this reduces to
\begin{equation}
V_{\rm REV}
=
\left[
L_{\rm REV}^{(3)}
\Delta x
\right]^3.
\label{eq:isotropic_three_dimensional_physical_volume}
\end{equation}

The spherical-shell average in
Eq.~\eqref{eq:three_dimensional_shell_average} defines an
orientation-averaged structural REV. When a directional structural REV is
required, reciprocal space can instead be partitioned into prescribed
directional sectors \(\mathcal D_\alpha\). Defining a separate residual
\(\eta_{\widehat C,\alpha}^{(3)}(L)\) within each sector leads to the
directional convergence condition
\begin{equation}
\max_\alpha
\eta_{\widehat C,\alpha}^{(3)}(L)
\le
\tau_{\widehat C}^{(3)}.
\label{eq:anisotropic_three_dimensional_spectral_rule}
\end{equation}
This optional directional formulation prevents spherical averaging from
concealing unconverged anisotropic correlations.

The observation size \(L\) is an external scale parameter and not a fourth
spatial coordinate. The three-dimensional analysis therefore consists of a
one-parameter family of voxelized subvolumes, rather than a stationary field
defined on a four-dimensional surface.

For volumetric physical validation, the structurally predicted support may be
transferred without property-specific recalibration to voxelized constitutive
fields. Three-dimensional thermal homogenization then yields an apparent
\(3\times3\) conductivity tensor, while elastic homogenization yields the full
three-dimensional apparent stiffness tensor. Stabilization of these independent
volumetric responses near the structurally predicted support would constitute
the direct three-dimensional counterpart of the physical validation performed
here. This validation is not required for the mathematical dimensional
generalization established above and is not claimed for the present
two-dimensional dataset.

%